\documentclass[aps,prb,twocolumn,superscriptaddress]{revtex4-2}

\usepackage{graphicx} 
\usepackage{graphics}
\usepackage{epsfig}
\usepackage{appendix}
\usepackage{amsmath, amssymb, braket}
\usepackage{dirtytalk}
\usepackage{hyperref, listings}
\usepackage{bbold}
\usepackage{esvect, cancel}
\usepackage{svg}
\usepackage{comment}
\usepackage{float}

\begin{document}

\title{Cavity Mediated Two-Qubit Gate: Tuning to Optimal Performance with NISQ Era Quantum Simulations}

\author{Shreekanth S. Yuvarajan} 
\email{shreekan@buffalo.edu}
\affiliation{Department of Physics, University at Buffalo SUNY, Buffalo, New York 14260, USA}
\author{Vincent Iglesias-Cardinale}
\affiliation{Department of Physics, University at Buffalo SUNY, Buffalo, New York 14260, USA}
\author{David Hucul}
\affiliation{United States Air Force Research Laboratory, Rome, New York, 13441, USA}
\author{Herbert F. Fotso}
\email{hffotso@buffalo.edu}
\affiliation{Department of Physics, University at Buffalo SUNY, Buffalo, New York 14260, USA}

\begin{abstract}
\noindent A variety of photon-mediated operations are critical to the realization of scalable quantum information processing platforms and their accurate characterization is essential for the identification of optimal regimes and their experimental realizations. 
Such light-matter interactions are often studied with a broad variety of analytical and computational methods that are constrained by approximation techniques or by computational scaling. Quantum processors present a new avenue to address these challenges.  We consider the case of cavity mediated two-qubit gates. To investigate quantum state transfer between the qubits, we implement simulations with quantum circuits that are able to reliably track the dynamics of the system. Our quantum algorithm, compatible with NISQ (Noisy Intermediate Scale Quantum) era systems, allows us to map out the fidelity of the state transfer operation between qubits as a function of a broad range of system parameters including the respective detunings between the qubits and the cavity, the damping factor of the cavity, and the respective couplings between the qubits and the cavity. The algorithm provides a robust and intuitive solution, alongside a satisfactory agreement with analytical solutions or classical simulation algorithms in their respective regimes of validity. It allows us to identify under-explored regimes of optimal performance, relevant for heterogeneous quantum platforms, where the two-qubit gate can be rather effective between far-detuned qubits that are neither resonant with each other nor with the cavity. Besides its present application, the method introduced in the current paper can be efficiently used in otherwise untractable variations of the model and in various efforts to simulate and optimize photon-mediated two-qubit gates and other relevant operations in quantum information processing.
\end{abstract}

\maketitle

\section{Introduction}
\label{sec:Introduction}

\noindent The optimization of light-matter interfaces arising in a variety of operations is key to the realization of scalable quantum platforms~\cite{Kimble_QtumInternet2008, DLCZ, ChildressRepeater, AharonovichEnglundToth_NatPhot, AwschalomHansonZhou_NatPhot, SchoelkopfGirvin_Nature2008, Haroche2006_book}. For instance, effective coupling enabling the implementation of quantum gates can be achieved between pairs or ensembles of qubits placed in a cavity~\cite{HagleyHaroche_PRL1997, FreyWallraff_PRL2012, EnglundLukin_NanoLett2010, ZhengGuo_PRL2000, YangHan_PRL2004, RosseauByrnes_PRA2014}. However, the efficiency of such light-mediated operations will depend on specific parameters of the emitters and of the cavity, and their precise characterization is essential for effective experimental realizations. Such accurate characterization of light-matter interactions in various quantum information processing (QIP) applications plays an important role in guiding technological advancements and in understanding intriguing emergent behaviors. Analytical solutions of suitable models are often constrained by various unavoidable approximations that are made to keep the problem tractable~\cite{ MarkovApprox_WodEberl, AckerhaltEberly, RF_KimbleMandel77, Milonni}. Such approximations may for instance involve integrating out the bosonic modes and solving for the density matrix operator of the qubits~\cite{RF_Mollow_PhysRev1969, Cohen_Tannoudji_Book1992, Loudon_book1983}.
They are often constrained in their validity to specific regions of parameter space and can easily lead to misleading spurious results. Alternatively, computational methods can be employed to tackle the problem. These numerical solutions, in addition to suffering from typical scaling issues when addressing quantum systems with classical computing algorithms, often involve an additional layer of abstraction between the model of the physical system and the classical algorithm.\\
Nascent quantum computers have brought about new possibilities for the simulation of quantum systems~\cite{Preskill2018quantumcomputingin, Fauseweh2024, DaleyZoller_Nature2022, BhartiAspuru-Guzik_RMP2022, IppolitiKhemani_prx2021, FotsoSPIE2025}. Although quantum processors are still in their infancy, quantum algorithms, even when solved on quantum simulators, offer a pathway to simulate quantum systems through a mapping that affords the user the capacity to directly track the dynamics of a quantum system. \\ 
Two-qubit gates can be performed through light-mediated interaction between qubits placed in a cavity e.g. between quantum dots, superconducting qubits or atoms in a cavity~\cite{HagleyHaroche_PRL1997, FreyWallraff_PRL2012, EnglundLukin_NanoLett2010, ZhengGuo_PRL2000, YangHan_PRL2004, RosseauByrnes_PRA2014}. To characterize and optimize these QIP operations, a variety of analytical and computational methods have been previously used. Furthermore, a number of initiatives have explored various methods to optimize this process using analytical and numerical approaches~\cite{WarrenBarnesEconomouPRB2021, WarrenBarnesEconomouPRB2019, SrinivasaTaylorPettaPRX2024, BenitoPettaBurkardPRB2019, Fotso_TPI_PRB_2019, FotsoEtal_PRL2016, Fotso_noisyTLS2022, FotsoOtherPulsesJPhysB2018, Beaudoin_2017}.

\noindent In this paper, we study the efficiency of this photon-mediated operation by implementing quantum circuits that are compatible with NISQ (Noisy Intermediate-Scale Quantum) era systems. The algorithm solves the quantum optics model describing the dynamics of the system, and allows us to reliably characterize the validity of conventional approximations that often inhibit analytical/numerical solutions. We focus on the case of a quantum state transfer and our approach allows us to assess the fidelity of the operation between two qubits and its dependence on various system parameters such as the emission frequencies of the qubits, the cavity frequency, the respective qubit couplings to the cavity, as well as the damping factor of the cavity. We examine the effect of the rotating wave approximation on the state transfer operation. We are also able to access the dispersive regime without any need for a perturbative solution such as the Schrieffer-Wolff transformation.\\

An important consideration is that computational scaling remains an important hurdle in the adoption of quantum algorithm for many-particle systems. 
In particular, for systems involving bosonic degrees of freedom, this problem is exacerbated by the possible large number of excitations in one mode. 
However, for the problem at hand, as is the case in various light-matter processes in QIP, this concern is alleviated by the fact that the system can be reasonably considered to be isolated during the timescale of the simulation. The number of excitations in the entire system is limited and is capped by the preparation of the initial state (e.g. a single excitation, for one qubit in the excited state, the other in the ground state and an empty cavity). In this way, the simulation can be performed with only a handful of qubits (3 qubits for two qubits in an ideal cavity, 4 qubits for two qubits in a cavity with a finite damping factor). We note that the Hamiltonian we simulate in the present paper allows for transient non-energy-conserving state transitions which include higher excited states of the cavity, suggesting that we may need more than 1 qubit to represent the multi-level cavity. However, as discussed in Appendix \ref{sec:Appendix_Multilevel_Cavity}, such transitions are negligible when the parameters of the system are within the regime of validity of the Rotating Wave Approximation (RWA), which is the case for most of our discussion when we benchmark our results against analytical results requiring the RWA. \\
It is clear from the problem size that this specific situation can be handled by exact diagonalization on a classical computer. However, in addition to its application presented in the current paper,  the method utilized for the present studies could be readily generalized to larger problem sizes including variations of cavity-mediated two-qubit gates with many radiation modes in the cavity. Also, the method is rather robust and intuitive in its implementation.  For this reason, it can be efficiently extended to broader efforts to simulate and optimize photon-mediated two-qubit gates and other relevant operations in quantum information processing that use light-matter interfaces~\cite{WarrenBarnesEconomouPRB2021, WarrenBarnesEconomouPRB2019, SrinivasaTaylorPettaPRX2024, BenitoPettaBurkardPRB2019, Fotso_TPI_PRB_2019, FotsoEtal_PRL2016, Fotso_noisyTLS2022, FotsoOtherPulsesJPhysB2018, Beaudoin_2017}.

\noindent The rest of the paper is structured as follows. In section~\ref{sec:Model}, we describe the model describing the system of interest and the associated Hamiltonian. In section~\ref{sec:Methods} we present the procedure that we follow, starting with a mapping of the Tavis-Cummings Hamiltonian onto a Hamiltonian of qubits, followed by the Suzuki-Trotter decomposition for the time-evolution of the quantum state under the ``qubitized" Hamiltonian, and then the quantum circuit that is used for this time evolution. We also discuss the systematic errors due to the Suzuki-Trotter decomposition of our Hamiltonian. In section~\ref{sec:polarizedState}, we discuss our analysis for the transfer of the fully polarized state and present the effects of a finite damping parameter on the cavity. In section~\ref{sec:DifferentCouplings}, we examine the state transfer operation when the two emitters have different coupling strengths to the cavity. In section~\ref{sec:SuperpositionState} we present our analysis of the state transfer for the superposition state. In section~\ref{sec:RWA}, we discuss the rotating wave approximation and its regimes of validity. We end with our conclusions in section \ref{sec:conclusion}. The analytical solutions for a few special situations are shown in the appendix.

\section{Model}
\label{sec:Model}

\begin{figure}[t]
 \includegraphics[width=8.40cm]{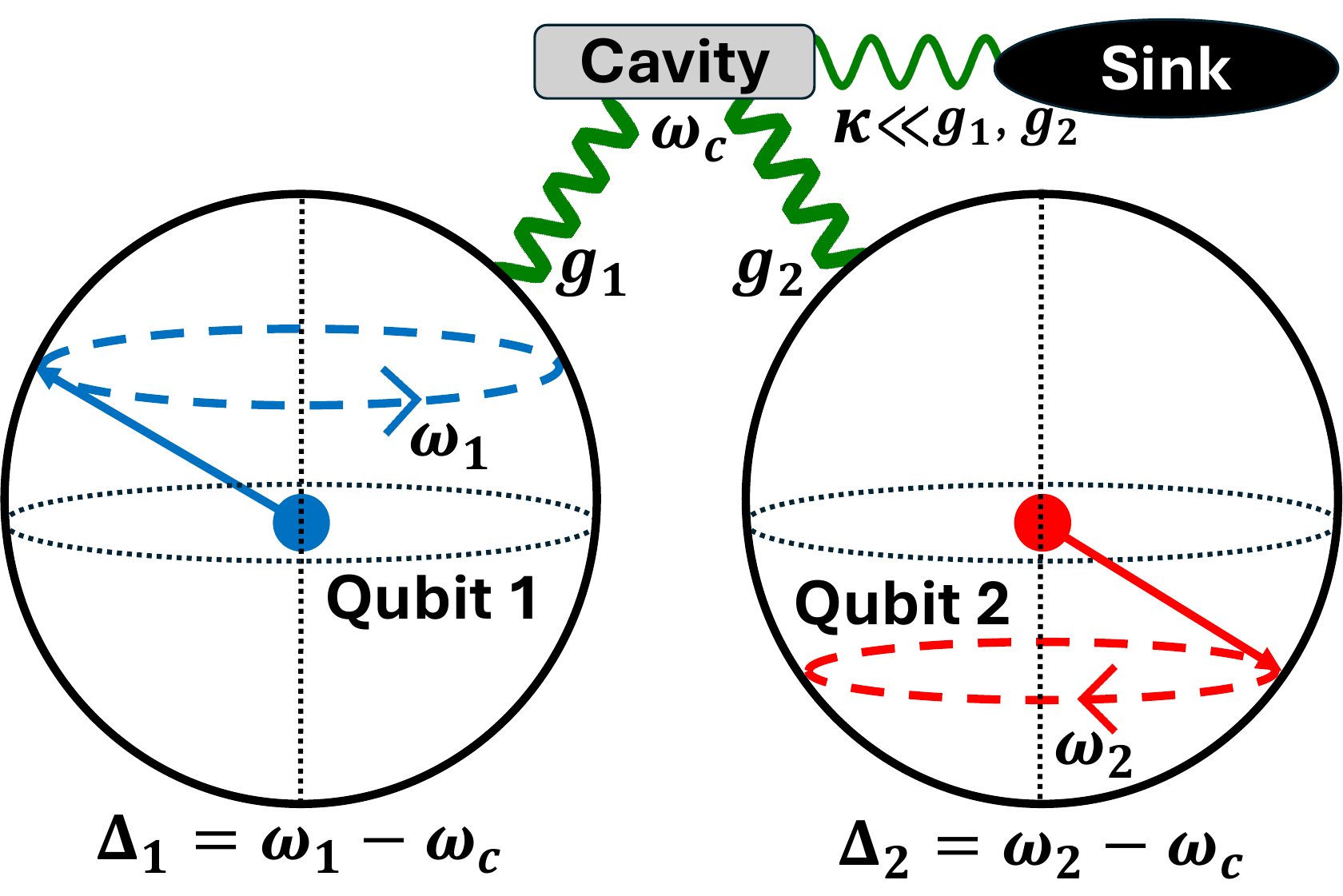}
    \caption{Schematic description of a system of two qubits  coupled through a cavity. \textit{Qubit 1} and \textit{Qubit 2} operate at frequencies $\omega_1 = \omega_c + \Delta_1 $ and $\omega_2 = \omega_c + \Delta_2 $ respectively and are coupled to the cavity with respective coupling strengths $g_1$ and $g_2$. $\Delta_1$ and $\Delta_2$ are the detunings of \textit{Qubit 1} and \textit{Qubit 2} respectively with the cavity that has a damping factor $\kappa$ and is at frequency $\omega_c$. The damping process is represented by the coupling of the cavity with a sink.
}
    \label{fig:system}
\end{figure}

\noindent We study the system that can be pictured schematically by Fig.~\ref{fig:system}. The figure describes a system of two qubits  coupled through a cavity.  While this iteration of the two-qubit gate can be treated with a classical computer, we use our quantum algorithm that would be readily applicable to more complex cases beyond the capacity of classical hardware. \textit{Qubit 1} and \textit{Qubit 2} operate at frequencies $\omega_1 = \omega_c + \Delta_1 $ and $\omega_2 = \omega_c + \Delta_2 $ respectively and are coupled to the cavity with respective coupling strengths $g_1$ and $g_2$. $\Delta_1$ and $\Delta_2$ are, respectively, the detunings of \textit{Qubit 1} and \textit{Qubit 2} with the cavity that has a damping factor $\kappa$ and is at frequency $\omega_c$. The damping process is represented by the coupling of the cavity with a sink. Such a system serves as a model for studying state transfer dynamics between the qubits for a variety of conditions.

\noindent In what follows we will first consider the case of equal coupling strengths between the qubits and the cavity i.e $g_1 = g_2$ and discuss it extensively. We will later consider the case where the couplings are different in section~\ref{sec:DifferentCouplings}.
We will also focus the initial discussion on the case of an ideal cavity ($\kappa = 0$) before mentioning the case of finite damping in section~\ref{subsec:dampedCavity}. \\

\noindent For a perfect cavity, $\kappa = 0$, the system can be described by the Tavis-Cummings Hamiltonian~\cite{TavisCummings1968} that extends the Jaynes-Cummings Hamiltonian~\cite{JaynesCummings1963} to an ensemble, here, a pair of two-level systems corresponding to our qubits, placed inside a cavity. Its Hamiltonian is:
\begin{eqnarray}
H &=& \frac{\hbar \omega_1}{2}\sigma^z_1 + \frac{\hbar \omega_2}{2}\sigma^z_2 + \hbar \omega_c a^{\dagger}_c a_c \nonumber \\
&\quad& + \hbar g_1 (a_c^\dagger \sigma^-_1 + a_c \sigma^+_1 + a_c^\dagger \sigma^+_1 + a_c \sigma^-_1) \nonumber \\ 
&\quad& + \hbar g_2 (a_c^\dagger \sigma^-_2 + a_c \sigma^+_2 + a_c^\dagger \sigma^+_2 + a_c \sigma^-_2)
\label{eq:Hamiltonian}
\end{eqnarray}

\noindent Here, $\omega_1$, $\omega_2$ and $\omega_c$ are, respectively, the frequencies of \textit{Qubit 1}, \textit{Qubit 2}, and the cavity. $a^{\dagger}_c$ and $a_c$ are respectively the creation and annihilation operators for the radiation mode at the frequency $\omega_c$. $\sigma^z$, $\sigma^-_i = |g_i\rangle\langle e_i |$ and $\sigma^+=(\sigma^-)^\dagger$ are, respectively, the $z$-axis Pauli matrix, the lowering, and the raising operators for the two-level system, with $|g_i\rangle$ and $ |e_i\rangle$ the ground and the excited states of \textit{Qubit i}.

\noindent In general, the state of the system at time $t$ can be written as:
\begin{equation}
    |\Psi(t)\rangle = \sum_l \alpha_l(t)|\Phi^l_1\rangle \otimes |\Phi^l_c\rangle \otimes|\Phi^l_2\rangle. 
\end{equation}
Where $|\Phi^l_i\rangle$ with $i = 1,\; 2$ is the state of \textit{Qubit i},  $|\Phi^l_c\rangle$ is the state of the cavity and  $\alpha_l(t)$ is a complex number. We have, $|\Phi^l_i\rangle = | e_i \rangle$ or $| g_i \rangle$. 
$|\Phi^l_c\rangle = | 0_c \rangle$ or $| 1_c \rangle$, with $ | 0_c \rangle$ and $ | 1_c \rangle$ the ground and the first excited state of the cavity.  Since each of the $|\Phi^l_j\rangle$ with $j = 1,\; 2,\; c$ lives in a two-dimensional space, the state of the system exists in an eight-dimensional space. Thus, the problem is not generally tractable analytically. There are, however, analytical solutions for special parameter sets, some of which we discuss later in the paper.\\
The goal of the quantum state transfer is to achieve the transfer of the state $| \Phi_1 \rangle$ onto \textit{Qubit 2}. Thus, the fidelity of the operation after time $t$ is $F(t) = |\langle \Psi_{target}|\Psi(t)\rangle|^2$ with $|\Psi(0) \rangle=|\Phi_1 0_c g_2\rangle$ and $|\Psi_{target} \rangle=| g_1 0_c \Phi_2 \rangle$. 
We use this quantity to examine the quantum state transfer operation, when the system is initialized by setting \textit{Qubit 1} in state $| \Phi_1 (0) \rangle$ while \textit{Qubit 2} is in state $| \Phi_2(0) \rangle$ and the cavity is kept in its ground state. In all solutions presented in this work, $| \Phi_2(0) \rangle$ = $| g \rangle$.
We will show that although it is not in general possible to compute this quantity for all system parameters analytically, our quantum circuits allow for an extensive examination of the operation throughout the parameter space.

\section{Methods}
\label{sec:Methods}

\noindent The overall solution is summarized in the flowchart shown in Fig.~\ref{fig:flowchart}. The left-hand side of the figure depicts the analytical procedure of ``qubitizing" the Hamiltonian, i.e. mapping the Tavis-Cummings Hamiltonian on to a Hamiltonian with only qubits. This analytical component also includes the Suzuki-Trotter decomposition. The right-hand side of the figure highlights the steps of the quantum simulation to evolve the quantum state.

\subsection{From Tavis-Cummings to qubit Hamiltonian}

\noindent Beginning with the Hamiltonian in Eq.~\ref{eq:Hamiltonian} and setting $\hbar=1$, we apply the transformation \cite{Nielsen_Chuang_2010}:
\begin{eqnarray}
\sigma^z_i &=& -Z_i \nonumber \\
\sigma^+_i &=& \dfrac{1}{2}(X_i - iY_i) \nonumber \\
\sigma^-_i &=& \dfrac{1}{2}(X_i + iY_i) \nonumber \\
a^\dagger_c &\equiv& \sigma^+_c = \dfrac{1}{2}(X_c - iY_c) \nonumber \\
a_c &\equiv& \sigma^-_c = \dfrac{1}{2}(X_c + iY_c).
\label{eq:HP_transformations}
\end{eqnarray}
Here, $X, \; Y, \; Z,  \mathrm{and} \; I $ are Pauli operators on the qubits and we use the subscript $c$ to identify the qubit corresponding to the cavity. This mapping of the cavity mode operators to Pauli operators is enabled by constraining the total excitation in the system to remain equal to $1$, which is reasonable when the system parameters are within the regime of validity of the RWA (see Appendices \ref{sec:Appendix_HPTransformation} and \ref{sec:Appendix_Multilevel_Cavity}, and Section \ref{sec:RWA}). 
Plugging these into the Hamiltonian in Eq.~\ref{eq:Hamiltonian}, we get the ``qubitized" Hamiltonian (see Appendix \ref{sec:Appendix_Qubitization}):

\begin{figure}[t]
    \centering
  \includegraphics[width=\columnwidth]{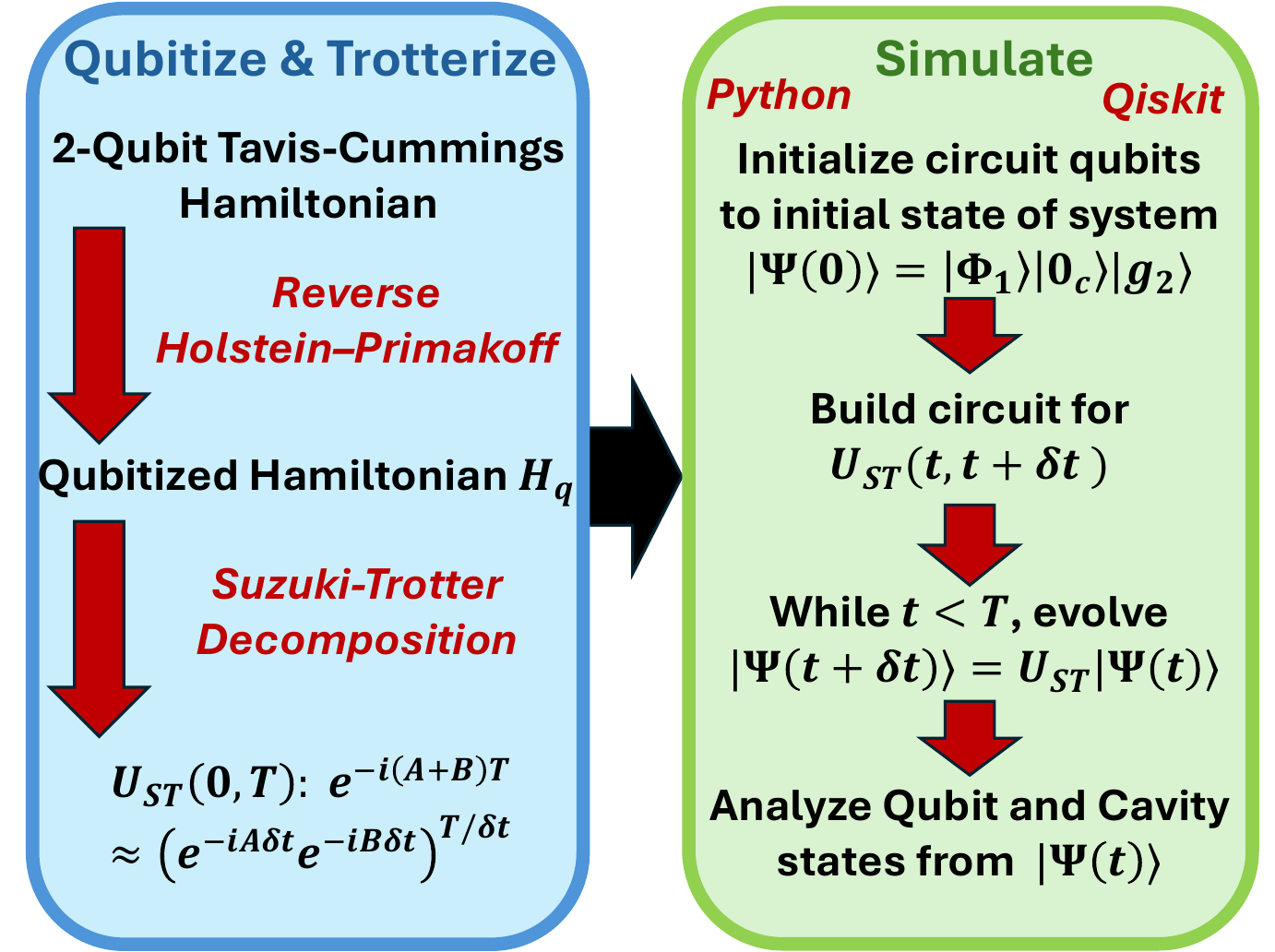}
    \caption{Illustration of the approach to solve the problem through the implementation of a quantum circuit. 
    The left-hand side depicts the analytical procedure of ``qubitizing" the Hamiltonian, i.e. mapping the Tavis-Cummings Hamiltonian on to a Hamiltonian with only qubits. This analytical component also includes the Suzuki-Trotter decomposition. The right-hand side highlights the steps of the quantum simulation to evolve the quantum state either on a quantum simulator or on a quantum processor.}
    \label{fig:flowchart}
\end{figure}

\begin{eqnarray}
H_q &=& -\dfrac{\omega_1}{2}Z_1 -\dfrac{\omega_2}{2}Z_2 
+ \dfrac{\omega_c}{2}(I_c - Z_c) \nonumber \\ 
&\quad& + g_1(X_c X_1) + g_2(X_c X_2).
\label{eq:qubitized_Hamiltonian}
\end{eqnarray}


\noindent Since we are interested in evolving the system in time, we use the Suzuki-Trotter decomposition to break the time-evolution operator into its action over finite time steps $\delta t$.
\begin{equation}
  U(0, t) \approx  \left( \mathrm{e}^{-\mathrm{i}H_0 \delta t} \mathrm{e}^{-\mathrm{i}H_{int} \delta t} \right)^{t/\delta t}
\end{equation}
so that
\begin{equation}
U(t, t+\delta t) \approx \mathrm{e}^{-\mathrm{i}H_0 \delta t} \mathrm{e}^{-\mathrm{i}H_{int} \delta t}  
\label{eq:trotterization}
\end{equation}
with
\begin{eqnarray}
H_0 &=& -\dfrac{\omega_1}{2}Z_1 -\dfrac{\omega_2}{2}Z_2 
+ \dfrac{\omega_c}{2}(I_c - Z_c) \\
H_{int} &=& g_1(X_c X_1) + g_2(X_c X_2)
\end{eqnarray}

\subsection{Quantum circuit and systematic error}
\noindent With the appropriate initial state, the quantum state of the system of qubits is evolved in time using the above mentioned Suzuki-Trotter decomposition that breaks the time evolution into terms that can be simulated with gates on appropriate qubits. The initialization and the time evolution are illustrated on the right-hand side of Fig.~\ref{fig:flowchart}. The number of time steps is chosen to minimize the systematic error while allowing simulations up to multiple relaxation times of the individual emitters. The quantum computation is performed with a Python code using the Qiskit library~\cite{Qiskit2024}.

\begin{figure}[t] 
    \centering  \includegraphics[width=\columnwidth]{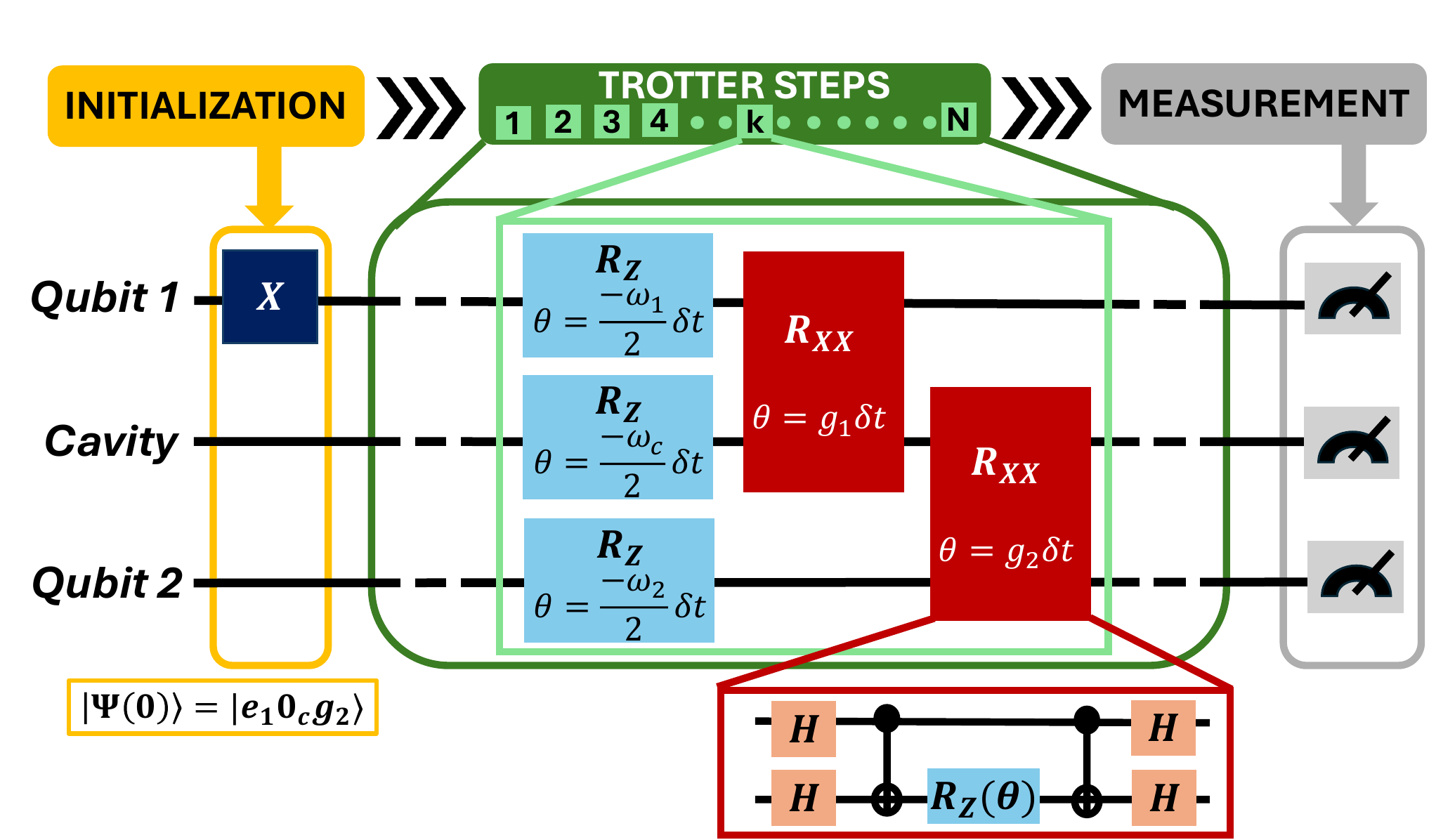}
    \caption{Quantum circuit for the time evolution of the initial state of a two-qubit system in a cavity under the Tavis-Cumings Hamiltonian. The circuit includes 3 qubits. The second qubit in the circuit represents the cavity while the first and the third represent \textit{Qubit 1} and \textit{Qubit 2} respectively, in our Tavis-Cummings Hamiltonian. 
    }
    \label{fig:quantumCircuit}
\end{figure}

The quantum circuit is shown schematically in Fig.~\ref{fig:quantumCircuit}. The figure shows 3 qubits, the second of which represents the cavity while the first and the third represent \textit{Qubit 1} and \textit{Qubit 2} respectively, in our Tavis-Cummings Hamiltonian qubitized into Eq.~\ref{eq:qubitized_Hamiltonian}.

In order to characterize the systematic error due to finite timestep discretization, we analytically estimate and analyze this error through the simulations on the quantum circuits of the Hamiltonian $H_q$. To help highlight key timescales involved in $H_q$, we transform it to the frame rotating at the cavity frequency(see Appendix \ref{sec:Appendix_Trotter_Error}):

\begin{eqnarray}
H_q^{cavity} &=& \sum_{i=1,2}   -\dfrac{\Delta_i}{2}Z_i + \dfrac{g_i}{2}(X_c X_i + Y_c Y_i) \nonumber \\ 
&\quad& \quad \quad + \dfrac{g_i}{2}cos(2 \omega_c t) (X_c X_i - Y_c Y_i) \nonumber \\
&\quad& \quad \quad + \dfrac{g_i}{2}sin(2 \omega_c t)(X_cY_i + iY_cX_i)
\label{Qubitized_Hamiltonian_Cavity_Frame}
\end{eqnarray}

\noindent By examining this form of the Hamiltonian, we can estimate that the leading order error in the time evolution operator $U(t, t+\delta t)$, due the Suzuki-Trotter satisfies (see Appendix \ref{sec:Appendix_Trotter_Error})

\begin{equation}
\epsilon \sim \delta t^2 \bigg( g_1 \Delta_1  + g_2 \Delta_2  + g_1 g_2 \bigg)
\end{equation}

\noindent For $N$ total time steps, an upper bound on the time step $\delta t$, for a desired total accumulated error $\epsilon_{tot}$ in a simulation from time $t_0=0$ to time $T=N\delta t$ is given by:

\begin{equation}
\delta t \sim \dfrac{\epsilon_{tot}}{T(g_1 \Delta_1  + g_2 \Delta_2  + g_1 g_2)}
\end{equation}

\noindent We benchmark our simulations by choosing a time step that produces an error in fidelity, $1-F \le 0.001$ for the corresponding complete state transfer operation (resonant or dispersive). This translates to a Trotter error $\epsilon_{tot} \equiv \sqrt{1-F} = 0.0316$ \cite{IkedaEtAl_PRR6_2024, GilchristEtAl_PRA71_2005}. 

We consider for further analysis of the systematic error, the case of emitters resonant with the cavity $\Delta_1=\Delta_2=0$ and where $g_1=g_2=g$. The transfer of the fully polarized state from \textit{Qubit 1} to \textit{Qubit 2} occurs after time $T=\pi /\sqrt{2}g$ and the desired fidelity requires:

\begin{equation}
\delta t \sim 0.015g^{-1}
\end{equation}

\begin{figure}[t] 
    \centering
  \includegraphics[width=\columnwidth]{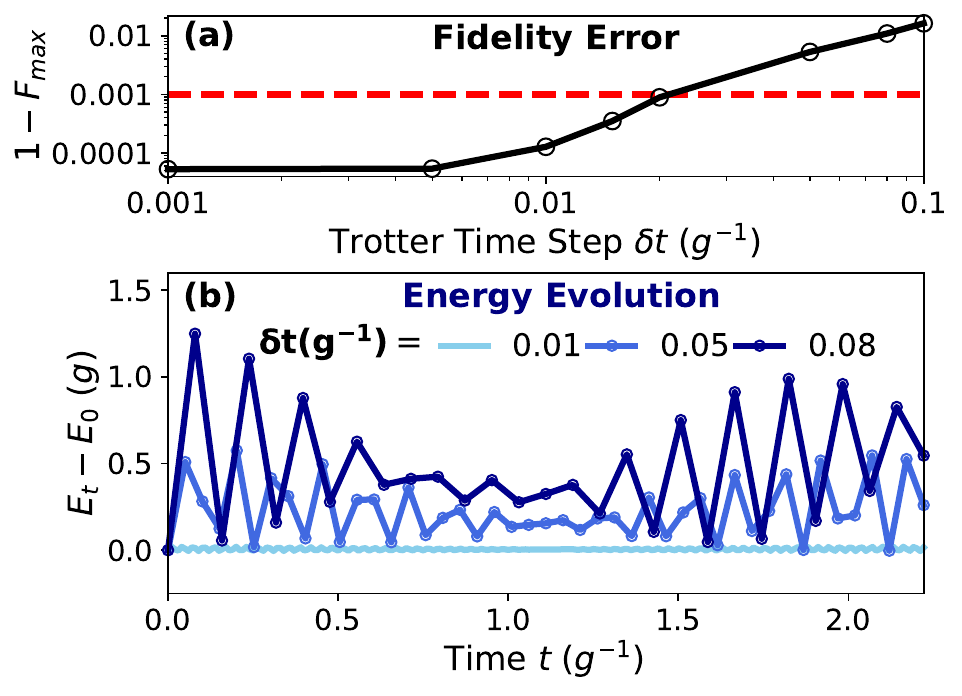}
    \caption{Illustration of the systematic error in the time evolution due to the Suzuki-Trotter decomposition. (a): Difference with ideal fidelity, $1-F_{max}$, for the transfer of the polarized state from \textit{Qubit 1} to \textit{Qubit 2} as a function of the Trotter time step $\delta t$. The dashed line represents the target of $1-F_{max}$ $\le 0.001$.    
(b): Evolution of the total energy in the system as a function of time for $\delta t = 0.01 \mathrm{g^{-1}}$, $\delta t = 0.05 \mathrm{g^{-1}}$ and $\delta t = 0.08 \mathrm{g^{-1}}$.
}
    \label{fig:timeStepBenchmarking}
\end{figure}

\noindent In Fig.~\ref{fig:timeStepBenchmarking}, we characterize the systematic error due to the Suzuki-Trotter discretization of the time axis by simulating the resonant transfer of the polarized state discussed above. The figure shows in panel (a) the fidelity error as a function of the time step $\delta t$. The dashed red line corresponds to our target error in the fidelity. Panel (b) shows the difference between the actual total energy in the system and the computed total energy, as a function of time for different choices of the time step. We see a clear suppression of the variation in the total energy of the system by reducing $\delta t$. We also note that the symmetry in the dynamics of the system is reflected in the energy deviation as a function of time, which is lowest when the excitation is most evenly distributed (at $t=T/2$), and highest when it is solely held in one of the qubits (at $t=0,T$).

In what follows, with the exception of section~\ref{sec:DifferentCouplings}, we mostly consider the case of $g_1=g_2=g$. We separately examine the case of the transfer of the fully polarized state and the case of the superposition state. All simulations are performed with $g=1$, thus setting our energy and time units by the coupling strength between the cavity and the qubits. Except for the simulation results in Fig. \ref{fig:RWAvsNRWA}(a), all simulations are performed with cavity frequency $\omega_c=100g$.

In order to characterize particular subsystems, and specific interactions between them, we construct the density matrix $\rho(t) = |\Psi(t)\rangle \langle\Psi(t)|$ for the global system state $|\Psi(t) \rangle$.

To obtain the reduced density matrices of \textit{Qubit 1}, \textit{Qubit 2} or the cavity, we trace out \textit{Qubit 2} and cavity, \textit{Qubit 1} and cavity, or \textit{Qubit 1} and \textit{Qubit 2} respectively. We can extract the effective coupling strength between a pair of subsystems (coherences) by tracing out the third subsystem to build a reduced density matrix ($4X4$) representing the pair. Such reduced density matrices are written as

\begin{eqnarray}
\rho_i(t) = Tr_{j,k}[\rho(t)] \\
\rho_{ij} (t) = Tr_k[\rho(t)]
\end{eqnarray}

\noindent where $(i,j,k) \in \{(1,c,2), (2,c,1), (c,1,2)\}$. From $\rho_i(t)$, one can get the populations and phases of the qubits and cavity as $|c_e|^2$ and $arg(c_e c_g^*)$ respectively. From $\rho_{ij}(t)$, one can get the coherence between subsystems $i$ and $j$ as $arg(c_{e0}c_{g1}^*)$ for $i,j=1(2),c$ and $arg(c_{eg}c_{ge}^*)$ for $i,j=1,2$.

\section{Transfer of the Polarized State}
\label{sec:polarizedState}

Except for select special cases, 
characterizing the state transfer between the two qubits is rather non-trivial in general. We first consider the case of a polarized state in \textit{Qubit 1} to be transferred onto \textit{Qubit 2}. \\

\subsection{Qubits resonant with the cavity} \noindent For the case of qubits that are resonant with the cavity, as we will discuss in section~\ref{sec:RWA}\textcolor{blue}{,} our simulations coincide with the rotating wave approximation (RWA) for which an analytical solution is achievable. Thus, we can use this RWA result to benchmark the quantum circuit results. 
Indeed, for $\Delta_1 = \Delta_2 = 0$, the RWA Hamiltonian in the frame rotating at the cavity frequency, and with $g_1 = g_2 = g$ is given by:
\begin{eqnarray}
H = g(a^\dagger_c \sigma^-_1 + a_c \sigma^+_1 + a^\dagger_c \sigma^-_2 + a_c \sigma^+_2).  
\label{eq:HamiltonianResonant}
\end{eqnarray}

\noindent For the system with a single excitation (only one of the qubits in the excited state or only the first excited state of the cavity occupied), the dynamics is restricted to the subspace spanned by the set of three states: $\{\ket{e_10_cg_2}, \ket{g_11_cg_2}, \ket{g_10_ce_2}\}$. With the initial state $\ket{\Psi(0)} = \ket{e_10_cg_2}$, the time-dependent state under the Hamiltonian in Eq.~\ref{eq:HamiltonianResonant} is (see Appendix \ref{sec:Appendix_ExactSolution}):
\begin{eqnarray}
\ket{\Psi(t)} &=&  \bigg( \dfrac{1 + cos \big(\sqrt{2}gt \big)}{2} \bigg) \ket{e_1 0_c g_2} \nonumber \\
&\quad& - \bigg( \dfrac{i sin \big(\sqrt{2}gt \big)}{\sqrt{2}} \bigg) \ket{g_1 1_c g_2} \nonumber \\
&\quad& + \bigg( \dfrac{cos \big(\sqrt{2}gt \big) - 1}{2} \bigg) \ket{g_1 0_c e_2}
\label{eqn:RWA_Evolution_Resonant}
\end{eqnarray}
From the above equation, it is easy to see that complete state transfer from \textit{Qubit 1} to \textit{Qubit 2} occurs at 
\begin{equation}
    t_f =\pi/\sqrt{2}g
\label{eq:RWA_TransferTimeResonant}
\end{equation}

\subsection{Dispersive regime}  
\noindent For qubits in the dispersive regime, the effective Hamiltonian in the RWA and in the frame rotating at the cavity frequency can be obtained through a Schrieffer-Wolf transformation to be~\cite{AmitDey_PRA2025}:
\begin{equation}
H_{eff} = \dfrac{g^2}{\Delta} (\sigma^z_1 + \sigma^z_2) + \dfrac{g^2}{\Delta}(\sigma^+_1 \sigma^-_2 + \sigma^-_1 \sigma^+_2)
\label{eq:EffectiveHamiltonianDispersive}
\end{equation}
This Hamiltonian in the dispersive regime eliminates the cavity and hence, the dynamics is constrained to the subspace spanned by the states: $\{\ket{e_1 g_2}, \ket{g_1 e_2}\}$. With the excitation initially in \textit{Qubit 1}, we get
\begin{equation}
   \ket{\Psi(t)} = \dfrac{1}{2}(1+e^{-2ig^2t/\Delta})\ket{e_1 g_2} + \dfrac{1}{2}(-1+e^{-2ig^2t/\Delta})\ket{g_1 e_2}
\label{eqn:RWA_Evolution_Dispersive}
\end{equation}
The time for complete state transfer ($\ket{e_1 g_2} \rightarrow \ket{g_1 e_2}$) in the dispersive regime is obtained as:
\begin{equation}
t^{'}_f = \pi \Delta / 2g^2
\label{eq:timeDispersiveHamiltonian}
\end{equation}

\noindent 
\subsection{General Solution}
\begin{figure}[t] 
    \centering
    \includegraphics[width=\columnwidth]{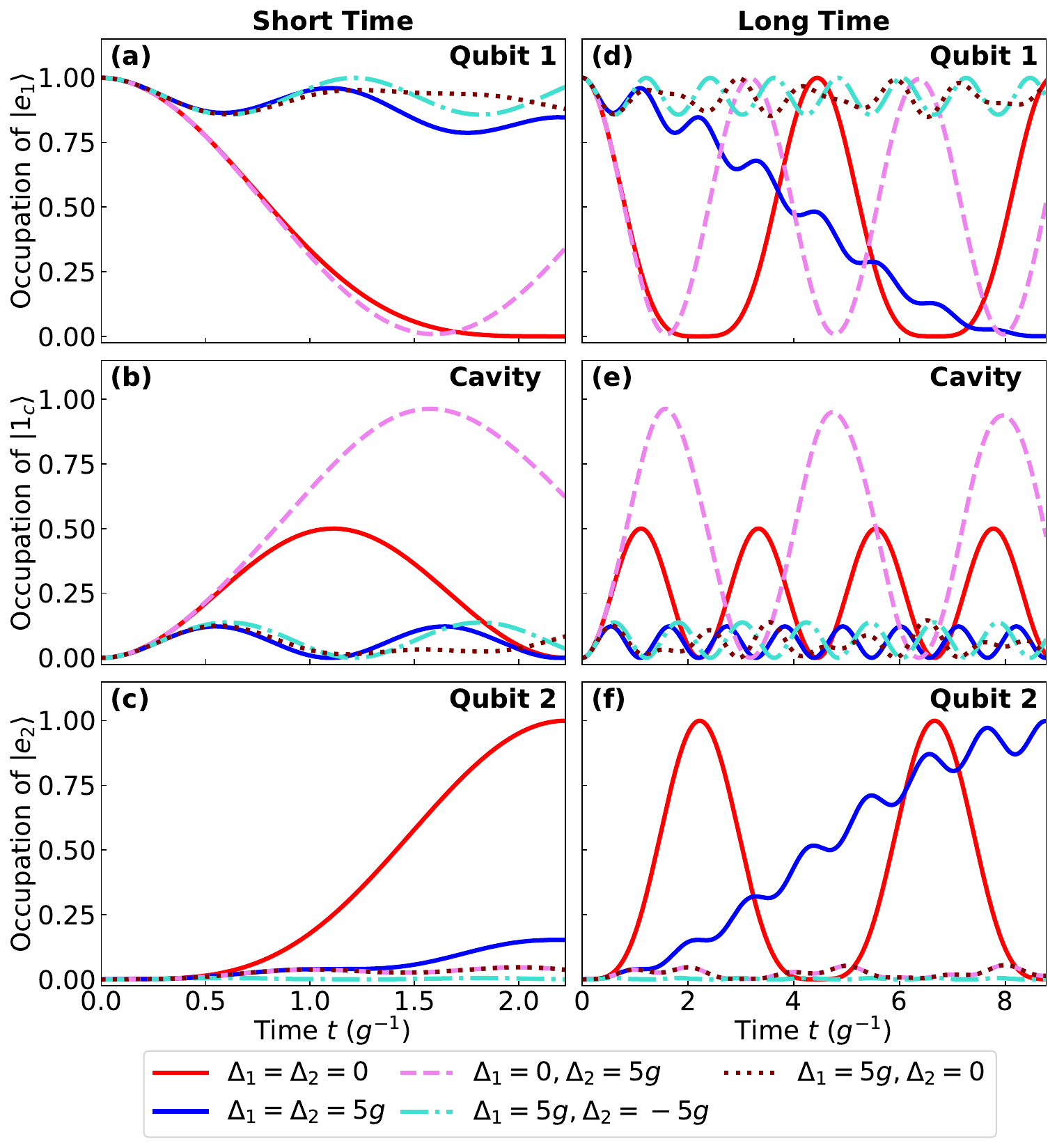}
    \caption{Evolution of the occupation of the qubits and the cavity excited states as a function of time. The system is initially in the state with \textit{Qubit 1} in its excited  state and \textit{Qubit 2} and the cavity in their ground  states. Panels (a), (b) and (c) show short time dynamics up to the time for first resonant state transfer $t_{f}$. Panels (d), (e) and (f) show long time dynamics up to the time for the first dispersive state transfer $t_{f}^{'}(\Delta=5g)$.}
    \label{fig:Polarized_State_Transfer}
\end{figure}

\noindent Fig.~\ref{fig:Polarized_State_Transfer} shows the solution with our quantum circuit for the general Hamiltonian (without the RWA) for different combinations of the detunings $\Delta_1$ and $\Delta_2$ selected to illustrate the variety of behaviors in the system. The figure presents the occupation of the excited states of \textit{Qubit 1} and \textit{Qubit 2} and of the first excited state of the cavity as a function of time. 
The left panels show the evolution up to a time $t_f$ at which the complete state transfer would occur in the resonant case. The right panels show the evolution of the system for longer times. The resonant case is represented by the solid red line, which at $t=t_f$, shows complete state transfer from \textit{Qubit 1} to \textit{Qubit 2} (with a global phase shift of $\pi$). Also, at $t=t_f/2$, the cavity reaches its maximum occupation, which is equal to half the total excitation in the system, in agreement with the analytical expression of Eq.~\ref{eqn:RWA_Evolution_Resonant}.

The detuning value of $\Delta = 5g$ is chosen to illustrate the dynamics as we approach the dispersive regime $\left( \Delta >> g \right)$. In general, it is observed that increased detuning suppresses excitation transfer. However, the symmetric detuning configuration, where the qubits are resonant with each other but detuned with the cavity (illustrated in the figure by $\Delta_1 = \Delta_2 = 5g$), shows a slow but near-complete transfer at a longer time, through effective qubit-qubit virtual coupling without ever significantly populating the excited state of the cavity.

The configuration in which \textit{Qubit 1} is resonant with the cavity and \textit{Qubit 2} is detuned (illustrated in the figure by $\Delta_1 = 0$ and $\Delta_2 = 5g$), shows coherent population oscillations between \textit{Qubit 1} and cavity, but only draws a feeble oscillatory response from \textit{Qubit 2}. The complementary configuration in which \textit{Qubit 1} is detuned from the cavity while \textit{Qubit 2} is resonant with the cavity (in the figure, $\Delta_1 = 5g$ and $\Delta_2 = 0$), shows weak oscillations in all subsystems. This is because even though \textit{Qubit 2} is resonant with the cavity, the system initially has the excitation in \textit{Qubit 1}, which is detuned from the cavity. The configuration in which both qubits are oppositely detuned (illustrated in the figure by $\Delta_1 = 5g$ and $\Delta_2 = -5g$) shows weak oscillations between \textit{Qubit 1} and cavity, and even weaker oscillations in \textit{Qubit 2} excited state occupation, due to the initial state of the system as mentioned above.

\subsection{Fidelity}

\begin{figure}[t] 
    \centering
    \includegraphics[width=\columnwidth]{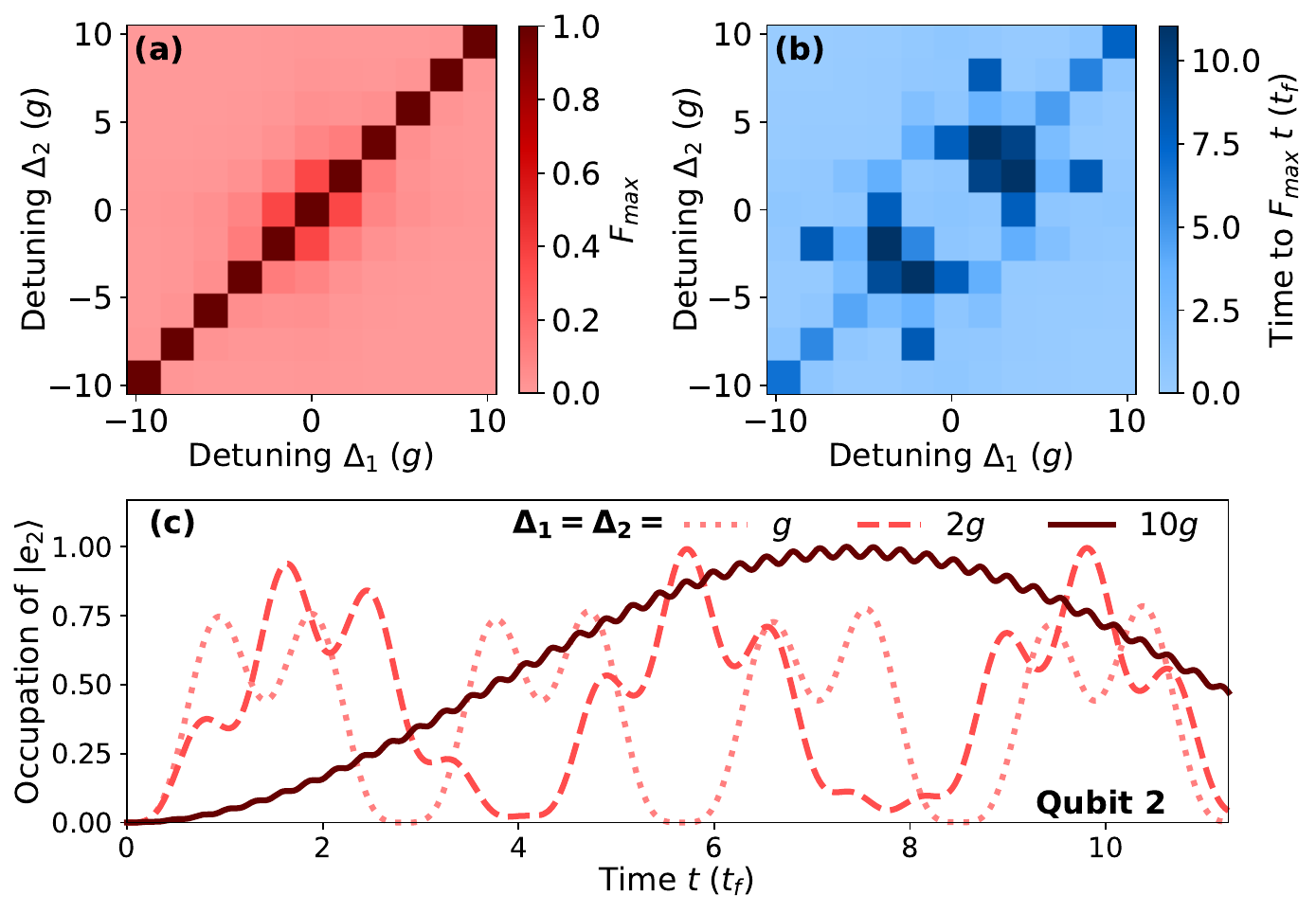}
    \caption{\textbf{(a)}: Heat map indicating the maximum achieved transfer fidelity $F_{max}$ as a function of detuning configurations. The resonant and symmetrically detuned configurations ($\Delta_1 = \Delta_2$) form the diagonal in the fidelity heat map and show complete state transfer, through different mechanisms. Maximum fidelity rapidly drops in value with increased detuning asymmetry. \textbf{(b)}: Heat map illustration of the time taken to achieve $F_{max}$ as a function of the detuning configurations. The resonant configuration achieves complete transfer early (at $t = t_{f}$). The dispersive configurations ($\Delta_1=\Delta_2 \gtrsim 5g$) achieve complete transfer over a longer time $t_{f}^{'}(\Delta)$. The intermediate configurations ($\Delta_1=\Delta_2 < 5g)$) achieve transfer through a growing envelope like phenomenon over much longer times.  
    Note the subtle asymmetry in darkness along the diagonal, between positive and negative equal detuning configurations (an effect missed by the RWA as discussed in section ~\ref{sec:RWA}).\textbf{(c)}: Occupation of the excited state of \textit{Qubit 2} as a function of time for different regimes of equal detuning configurations highlighting slow  dispersive state transfer and an even slower intermediate state transfer. 
    }
    \label{fig:Fmax_Polarized}
\end{figure}

\noindent To characterize the quantum state transfer operation  at time $t$, we use the fidelity defined by:
\begin{equation}
    F(t) = \left| \langle \Psi_{target} | \Psi(t) \rangle \right|^2
    \label{eq:fidelity}
\end{equation}
with $|\Psi(0) \rangle =|e_1 0_c g_2\rangle$ and $|\Psi_{target} \rangle =| g_1 0_c e_2 \rangle$.
To analyze the dependence of the efficiency on the parameters of the system, we solve the problem with our quantum circuit for a broad set of detuning values, evolving the system up to long times, and then extracting the maximum achieved fidelity. A figure of merit for the state transfer operation is then the maximum fidelity achieved, as well as the time to achieve this value.
Figure~\ref{fig:Fmax_Polarized} presents through a heatmap the maximum fidelity in panel (a), and the time to achieve this maximum fidelity in panel (b), as a function of the detunings $\Delta_1$ and $\Delta_2$ between \textit{Qubit 1} and the cavity and between \textit{Qubit 2} and the cavity, respectively. We observe that maximal fidelity is achieved for both qubits resonant with the cavity and with both qubits resonant with each other in the dispersive regime (diagonal values in Fig.~\ref{fig:Fmax_Polarized}(a)). The fidelity is lowered as the frequencies of \textit{Qubit 1} and \textit{Qubit 2} are tuned away from each other (off-diagonal values in Fig.~\ref{fig:Fmax_Polarized} (a)). In panel (b), we observe that the maximal fidelity achieved for $ \Delta_1 = \Delta_2 >> g$ (dispersive regime) is achieved over longer times (at $t_f^{'}(\Delta)$) than for both Qubits resonant with the cavity (at $t_f$). In the intermediate regime ($g <\Delta_1 = \Delta_2 <5g$), we observe maximal fidelity that is achieved over much longer times. 
Panel (c) shows further characterization of the long time dynamics through the occupation of the excited state of \textit{Qubit 2}. It clearly indicates that the maximum fidelity is achieved over a long timescale in the dispersive regime, and over an even longer timescale in the intermediate regime where complete state transfer is approached through a growing envelope of oscillations in the occupation of the excited state.

\subsection{State Transfer via a Damped Cavity}
\label{subsec:dampedCavity}

\noindent A realistic cavity is not perfect, and is subject to photon loss through cavity leakage. In this scenario, the pair of qubits and the cavity form an open system. Such an open system, with the inclusion of cavity loss can be modeled by introducing a photon loss mechanism represented by the addition to the Hamiltonian of the term $H_d = \omega_c a^\dagger_d a_d + \kappa a_c a^\dagger_d$, where $\kappa$ is the photon loss rate or the cavity damping rate, $a^\dagger_d$ and $a_d$ are respectively the creation and the annihilation operators for the bosonic mode of the sink to which the cavity is coupled. The overall damped Hamiltonian including the photon loss term becomes:
\begin{eqnarray}
H_{damped} &=& H_{TC} + H_d \nonumber \\
&=& \frac{\omega_1}{2}\sigma^z_1 + \frac{\omega_2}{2}\sigma^z_2 + \omega_c a^{\dagger}_c a_c \nonumber \\
&\quad& + g_1 (a_c^\dagger \sigma^-_1 + a_c \sigma^+_1 + a_c^\dagger \sigma^+_1 + a_c \sigma^-_1) \nonumber \\
&\quad& + g_2 (a_c^\dagger \sigma^-_2 + a_c \sigma^+_2 + a_c^\dagger \sigma^+_2 + a_c \sigma^-_2) \nonumber \\
&\quad& + \omega_c a^\dagger_d a_d + \kappa a_c a^\dagger_d
\label{eqn:Tavis_Cummings_H_lossy}
\end{eqnarray}
This Hamiltonian is not Hermitian and cannot be readily simulated using the Qiskit library. However, the equivalent effect of cavity loss can be simulated using an amplitude damping circuit \cite{gupta2020optimalquantumsimulationopen}. The circuit is based on a damping channel that uses Kraus operators, coupled to the cavity mode (see Appendix \ref{sec:Appendix_Damped_Cavity}). The arrangement uses an ancillary damping qubit, called the sink, which draws population out of the cavity at a rate $ \kappa$. The sink is reset to $|0\rangle$ before each Trotter step and measured after it, to ensure a ``one-way" flow of population from the cavity to the sink. The quantum circuit depicting damped Tavis-Cummings dynamics is shown in Appendix \ref{sec:Appendix_Damped_Cavity}.

\begin{figure}[t] 
    \centering
    \includegraphics[width=\columnwidth]{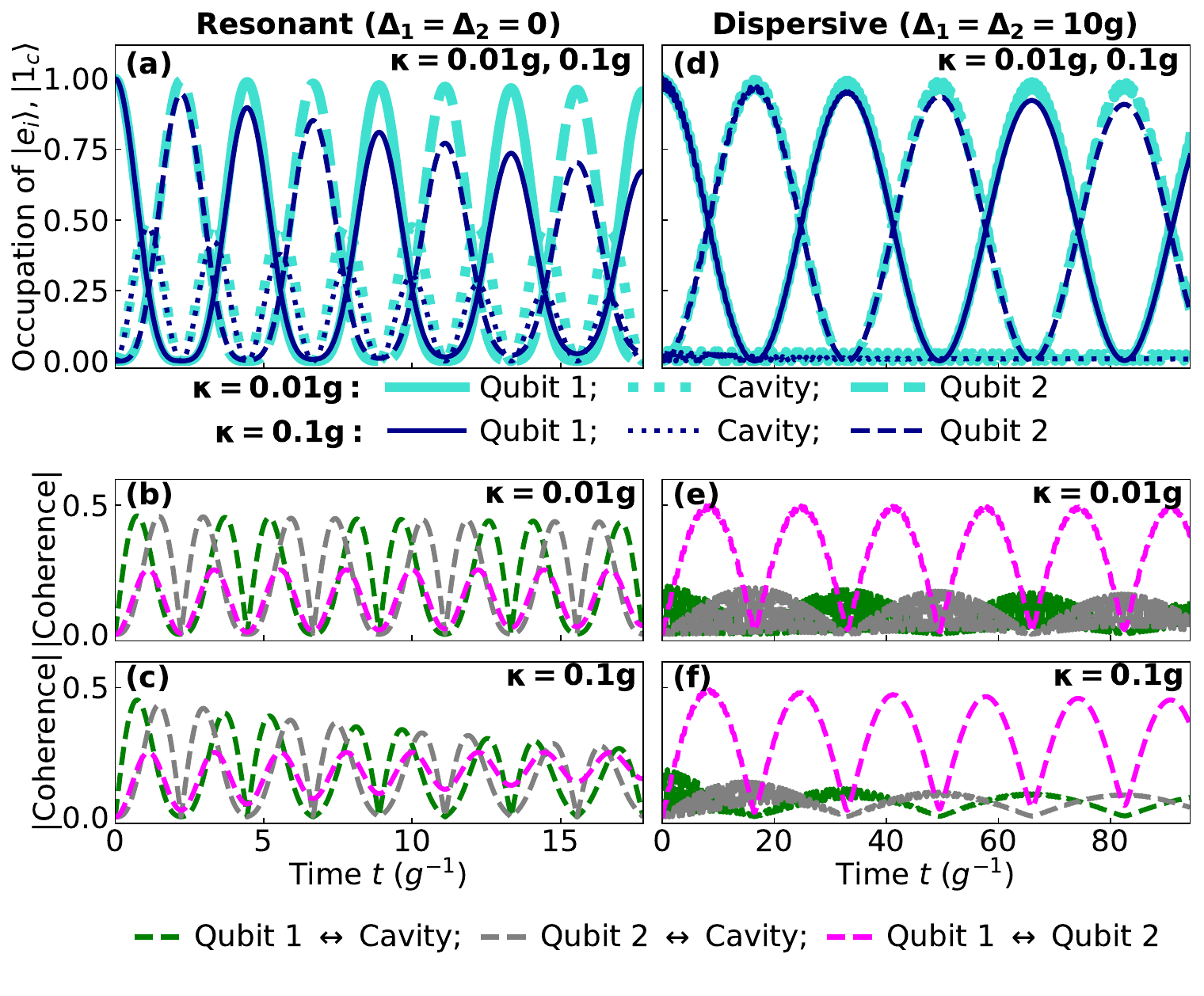}
    \caption{Evolution as a function of time of the excited state occupation of the qubits and the cavity for both qubits resonant with the cavity (a), and for both qubits resonant with each other in the dispersive regime (d). Evolution as a function of time of the coherences between pairs of subsystems, when the cavity is damped by a factor $\kappa = 0.01g$ (b) for the resonant case and (e) for the dispersive case, $\kappa = 0.1g$ (c) for the resonant case and (f) for the dispersive case.
    }
    \label{fig:Damped_Transfer}
\end{figure}

\noindent Figure~\ref{fig:Damped_Transfer} shows the dynamics of the system with qubits resonant with the cavity and with qubits in the dispersive regime for different values of the damping constant. Panel (a) shows the occupation as a function of time of the excited states of the qubits and the cavity for damping rate $\kappa = 0.01g$ and for $\kappa = 0.1g$ when the qubits are resonant with the cavity. Panel (d) shows the same quantities for qubits in the dispersive regime with $\Delta_1 = \Delta_2 = 10g$. We observe that in the resonant configuration, the system loses its overall excitation gradually at a rate proportional to the damping coefficient $\kappa$. However, for the dispersive configuration, for the same $\kappa$, we observe a much slower damping, owing to the transfer in this regime being less affected by the cavity loss as it is only virtually populated.
In the lower panels, the figure shows the coherences corresponding to the coupling of different pairs of subsystems. The figure clearly shows that the coherence between \textit{Qubit 1} and \textit{Qubit 2} (magenta dashed line in panels (b), (c), (e), and (f)) remains high in amplitude for the dispersive regime independently of the damping rate. Note also the suppression of the coherences between the qubits and the cavity (green and dark-gray dashed lines) in the dispersive regime.

\section{Transfer Between Qubits with Different Coupling Strengths to the Cavity}
\label{sec:DifferentCouplings}

\noindent Here, we simulate the transfer of the fully polarized state from \textit{Qubit 1} to \textit{Qubit 2} for the case of unequal coupling strengths between the qubits and cavity, for the resonant and dispersive configurations. \\

\subsection{Qubits resonant with cavity}

\noindent Here, the time evolution of the system in the single excitation subspace spanned by the set of three states: $\{\ket{e_10_cg_2}, \ket{g_11_cg_2}, \ket{g_10_ce_2}\}$,  is described by the quantum state (see Appendix \ref{sec:Appendix_ExactSolution})

\begin{eqnarray}
\ket{\Psi(t)} &=& \dfrac{g_1^2 cos \bigg(\sqrt{g_1^2 + g_2^2}t \bigg) + g_2^2}{g_1^2 + g_2^2} \ket{e_1 0_c g_2} \nonumber \\
&\quad& - \dfrac{i g_1 sin \bigg(\sqrt{g_1^2 + g_2^2}t \bigg)}{\sqrt{g_1^2 + g_2^2}} \ket{g_1 1_c g_2} \nonumber \\
&\quad& + \dfrac{g_1g_2 \bigg( cos \bigg(\sqrt{g_1^2 + g_2^2}t \bigg) - 1 \bigg)}{g_1^2 + g_2^2} \ket{g_1 0_c e_2}
\label{eqn:evolution_resonant_unequal_g_RWA}
\end{eqnarray}

\noindent One can see that the maximum occupation probability of the excited state of \textit{Qubit 2} is

\begin{equation}
max(P_2 )= \bigg( \dfrac{g_1g_2}{g_1^2 + g_2^2} \bigg( cos \bigg( \sqrt{g_1^2 + g_2^2}t \bigg) - 1 \bigg) \bigg)^2
\label{eqn:Qubit2_Max_Excitation_Probability}
\end{equation}

\noindent This is equal to $1$ only when $g_1=g_2$ and $t=\dfrac{\pi}{\sqrt{g_1^2 + g_2^2}} + 2n\pi$. Therefore, with unequal couplings, complete state transfer does not occur even if the qubits are resonant with the cavity.\\

\subsection{Dispersive regime}

\noindent For the dispersive configuration, with $\Delta_1 = \Delta_2$, the effective Hamiltonian in the RWA and in the frame rotating at the cavity frequency is obtained through a Schrieffer-Wolff transformation to be~\cite{AmitDey_PRA2025}:

\begin{equation}
H_{eff} = \sum_{i=1}^2\dfrac{g_i^2}{\Delta} \sigma^z_i + \sum_{i \ne j}\dfrac{g_i g_j}{2\Delta}(\sigma^+_i \sigma^-_j + \sigma^-_i \sigma^+_j)
\label{eq:EffectiveHamiltonianDispersive_UnequalCoupling}
\end{equation}

\noindent The time evolution of the system in the single excitation subspace now spanned by two states: $\{\ket{e_1 g_2}, \ket{g_1 e_2}\}$, is described by the quantum state \cite{AmitDey_PRA2025}

\begin{eqnarray}
\ket{\Psi(t)} &=& \bigg( 1 + \dfrac{g_2^2}{g_1^2 + g_2^2} \bigg( e^{-it(g_1^2 + g_2^2) / \Delta} - 1 \bigg) \bigg) \ket{e_1 g_2} \nonumber \\
&\quad& + \dfrac{g_1 g_2}{g_1^2 + g_2^2}\bigg( e^{-it(g_1^2 + g_2^2) / \Delta} - 1 \bigg)\ket{g_1 e_2}
\label{eqn:Evolution_Dispersive_unequal_g_RWA}
\end{eqnarray}

\noindent Here, the maximum occupation probability of the excited state of \textit{Qubit 2} is

\begin{equation}
max(P_2 )= \bigg( \dfrac{g_1g_2}{g_1^2 + g_2^2} \bigg( e^{-it(g_1^2 + g_2^2) / \Delta} - 1 \bigg) \bigg)^2
\label{eqn:Qubit2_Max_Excitation_Probability_Dispersive}
\end{equation}

\begin{figure}[t] 
    \centering   \includegraphics[width=\columnwidth]{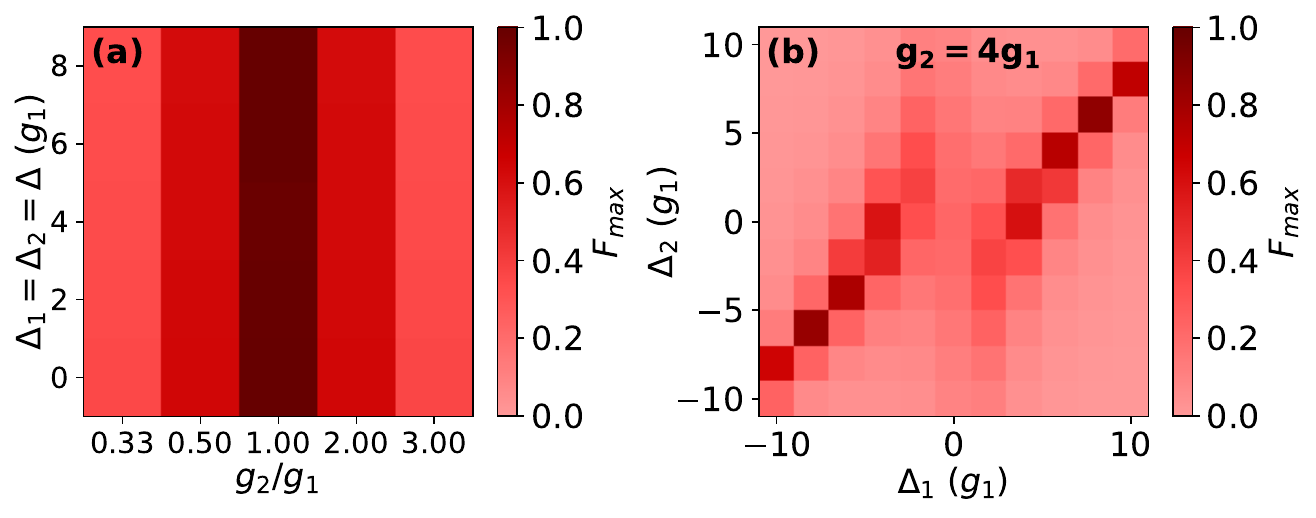}
    \caption{\textbf{(a)}: Heat map depicting the maximum achieved transfer fidelity $F_{max}$ for qubits resonant and equally detuned with the cavity as a function of the ratio of couplings $g_2/g_1$ for the transfer of the fully polarized state in \textit{Qubit 1}.  
    \textbf{(b)}: Heat map showing  $F_{max}$ as a function of qubit detunings $\Delta_1$ and $\Delta_2$ for $g_2/g_1=4$.}
    \label{fig:Unequal_Couplings}
\end{figure}

\noindent This is equal to $1$ only when $g_1=g_2$ and $t=\dfrac{\pi \Delta}{g_1^2 + g_2^2} + 2n\pi$. Therefore, with unequal couplings, complete state transfer does not occur in the dispersive regime. 
In Fig. ~\ref{fig:Unequal_Couplings}, panel (a) illustrates the drop in fidelity for the resonant state transfer of the polarized state as the ratio $g_2/g_1$ deviates from $1$.     $F_{max} \sim 1$ when $g_2/g_1=1$, and is $<1$ otherwise, in agreement with the analytical expressions in Eqs. \ref{eqn:Qubit2_Max_Excitation_Probability} and \ref{eqn:Qubit2_Max_Excitation_Probability_Dispersive}. Panel (b) shows the maximum state transfer fidelity as a function of qubit detunings when $g_2/g_1=4$. The state transfer is generally poor, including along the diagonal, representing the resonant and equally detuned ($\Delta_1 = \Delta_2$) qubit configurations, due to the inequality in coupling. However, some off-diagonal cells show a higher state transfer fidelity. These cells correspond to detuning combinations that compensate for the inequality in couplings by matching their Rabi frequencies with adequate detuning values to give $\Omega_i=\sqrt{g_i^2 + \Delta_i^2}$ such that $\Omega_1 \sim \Omega_2$.

\section{Transfer of Superposition State}
\label{sec:SuperpositionState}

\noindent The quantum circuits can be used to characterize the transfer of an arbitrary quantum state, including a superposition of the excited and ground state i.e. 
$|\Phi_1\rangle = \alpha_1 | e_1 \rangle + \beta_1 | g_1 \rangle$, with $\alpha_1$ and $\beta_1$ complex numbers.
For illustration, we focus here on the example of an equal superposition of the ground state and the excited state in Qubit 1 ($|\alpha_1| = |\beta_1| = 1/\sqrt{2}$). 

\subsection{Qubits Resonant with the Cavity}
\noindent For qubits that are resonant with the cavity, the model for the transfer of a superposition state requires that we expand the allowed subspace to the space spanned by the four-state set $\{ \ket{e_1 0_c g_2}, \ket{g_1 1_c g_2}, \ket{g_1 0_c e_2}, \ket{g_1 0_c g_2} \}$.

We note that the state $\ket{g_10_cg_2}$ is unaffected by the Tavis-Cummings Hamiltonian dynamics, rendering it a dark state. This, along with the linearity of the evolution operator $U(t)$, enables us to write the time-evolved state of the system as: 
\begin{eqnarray}
\ket{\Psi(t)} &=& \alpha\ket{\Psi_g(t)} + \beta\ket{\Psi_e(t)} \\
&=& \alpha\ket{\Psi_g(0)} + \beta \bigg(U(t)\ket{\Psi_e(0)} \bigg) \\
&=& \alpha\ket{g_1 0_c g_2} + \beta \bigg( U(t)\ket{e_1 0_c g_2} \bigg).
\end{eqnarray}

\noindent Where $\alpha$ and $\beta$ are complex numbers. Essentially, $|\Psi(t)\rangle$ for the superposition state only depends on the evolution of the excited component since the ground state component is preserved. Replacing $|\Psi_e(t) \rangle$ with Eq.~\ref{eqn:RWA_Evolution_Resonant}, we get 

\begin{eqnarray}
\ket{\Psi(t)} &=& \alpha \ket{g0g} + \beta \bigg[ \bigg( \dfrac{1 + cos \big(\sqrt{2}gt \big)}{2} \bigg) \ket{e_1 0_c g_2} \nonumber \\
&\quad& - \bigg( \dfrac{i sin \big(\sqrt{2}gt \big)}{\sqrt{2}} \bigg) \ket{g_1 1_c g_2} \nonumber \\
&\quad& + \bigg( \dfrac{cos \big(\sqrt{2}gt \big) - 1}{2} \bigg) \ket{g_1 0_c e_2} \bigg]
\label{eq:evolution_superposition_ideal}
\end{eqnarray}

\noindent We observe that the time for complete state transfer, $t_f = \pi/\sqrt{2}g$, is the same as that for the polarized case.

\subsection{Dispersive Regime}
\noindent Here, with the cavity dropping out of the subspace, the extended subspace accommodating the superposition state becomes the space spanned by the 3-ket set $\{ \ket{e_1 g_2}, \ket{g_1 e_2}, \ket{g_1 g_2} \}$. The evolution under the Tavis-Cummings Hamiltonian gives the time dependent state:

\begin{eqnarray}
   \ket{\Psi(t)} &=& \alpha \ket{g_1 g_2} + \beta \bigg( \dfrac{1}{2}(1+e^{-2ig^2t/\Delta})\ket{e_1 g_2} \nonumber \\
   &\quad& + \dfrac{1}{2}(-1+e^{-2ig^2t/\Delta})\ket{g_1 e_2} \bigg)
\label{eqn:RWA_Evolution_Dispersive_Superpositon}
\end{eqnarray}

We observe that the time for complete state transfer, $t^{'}_f = \pi \Delta / 2g^2$, is the same as that for the polarized state.

\subsection{General Solution}
\noindent In what follows, we use our quantum circuit to solve the problem for different combinations of detunings between the qubits and the cavity and we analyze both the short and the long-time dynamics.

\subsubsection{Short Time Dynamics}
\noindent We track the dynamics of the system up to a time corresponding to the complete state transfer ($t_f$) when the system is in a configuration with both qubits resonant with cavity. The simulation for this configuration agrees with the analytical expression in Eq.~\ref{eq:evolution_superposition_ideal}. Complete transfer of the superposition state from \textit{Qubit 1} to \textit{Qubit 2}, with the relative phase shifted by $\pi$, occurs at a time $t_f$. This phase shift is immediately observed at $t>0$, and is unperturbed for the duration of evolution.

\subsubsection{Long Time Dynamics}
\begin{figure}[t] 
    \centering
    \includegraphics[width=\columnwidth]{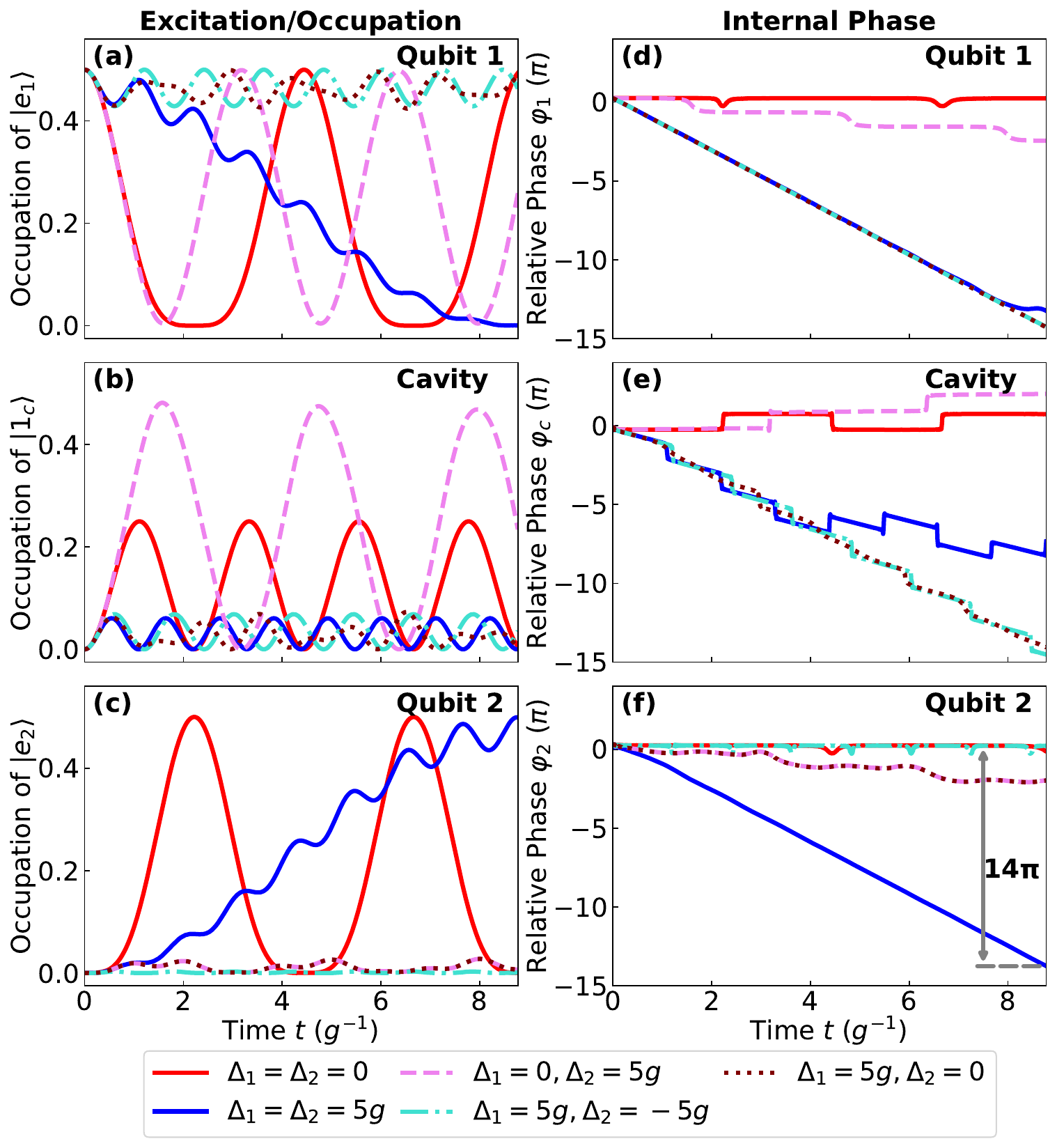}
    \caption{For 
 \textit{Qubit 1} initially in the equal superposition state, evolution of the excited state occupation of the qubits and the cavity (panels (a) through (c)), and their respective relative phases between the ground and the excited state (panels (d) through (f)), as a function of time. The system is tracked up to time $t_{f}^{'}(\Delta=5g)$ corresponding to the optimal transfer time in the dispersive regime. 
Phase data has been shifted to the cavity frame for illustration. The phase grows with a negative amplitude due to the mapping $\sigma_i^z=-Z_i$ as discussed in Section \ref{sec:Methods}.}
    \label{fig:superpositionTransferLong}
\end{figure}

\noindent Here, we explore the possibility of a slow but complete state transfer process by studying the long time dynamics of the system. This is presented in Figure~\ref{fig:superpositionTransferLong} where we show both the excited state occupation and the relative phase of \textit{Qubit 1}, the cavity, and \textit{Qubit} 2. 
Effective preservation ($\pi$ shifted) of the relative phase is seen in the case of qubits resonant with the cavity, and also in the equally detuned configuration. In the latter case, the relative phase in \textit{Qubit 2} appears to drift far away, but it is important to note that it has been unwrapped (along $2\pi$ cycles) for better plot readability. So, the relative phase drift, which is exactly $14 \pi$ (or seven $2 \pi$ cycles), is preserved.

\subsection{Fidelity}
\begin{figure}[t] 
    \centering
    \includegraphics[width=\columnwidth]{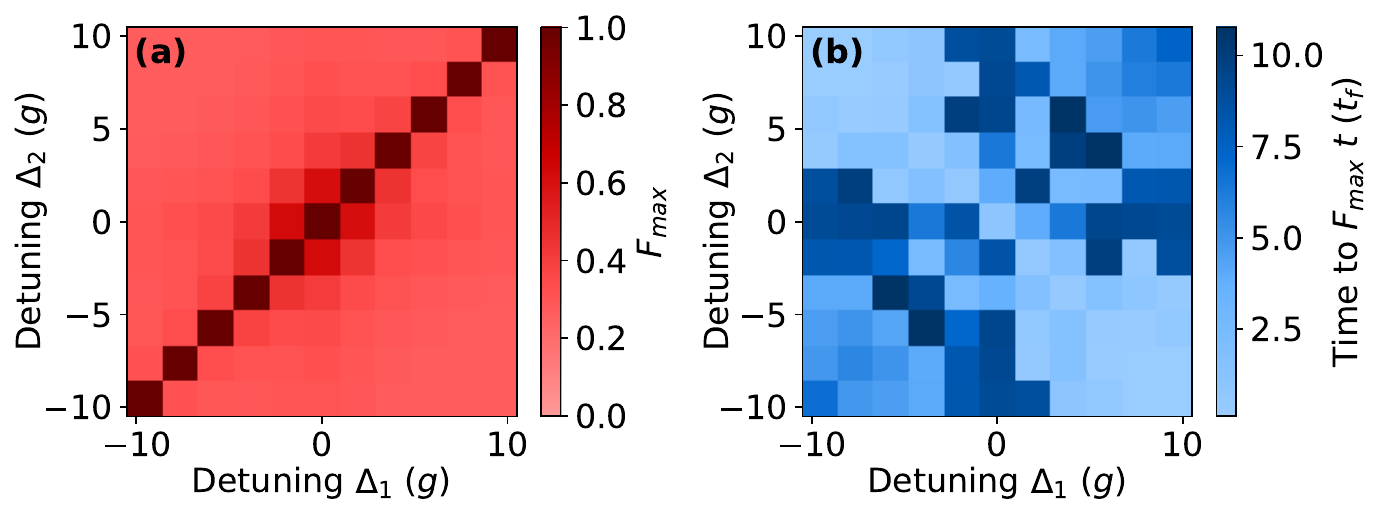}
    \caption{\textbf{(a)}: Heat map of the maximum achieved state transfer fidelity $F_{max}$ as a function of detuning configurations.   \textbf{(b)}: Heat map of the time to achieve $F_{max}$ as a function of detuning configurations.
    }
    \label{fig:Fmax_Superposition}
\end{figure}

\noindent We also analyze the fidelity of the state transfer operation in the case of this equal superposition state. As in the previous case, we compute the fidelity as a function of the detunings between the qubits and the cavity by evolving the system up to a long time and identifying the maximal fidelity achieved. We also record our time to achieve this maximum fidelity. The results are shown in Figs.~\ref{fig:Fmax_Superposition} (a) and (b). We note that the resonant and symmetrically detuned configurations ($\Delta_1 = \Delta_2$) form the diagonal and show complete state transfer, through the different mechanisms discussed above. Maximum fidelity rapidly drops with increasing detuning asymmetry. However, the smallest $F_{max}$ is higher than that of the transfer of the fully polarized state, an effect that is due to the presence of the passive ground state component $|g_1 0_c g_2 \rangle$ which is preserved in the evolution (within the regime of validity of the RWA, as discussed in section~\ref{sec:RWA}). Just as in polarized state transfer, the resonant configuration achieves maximal fidelity early (at $t_f$) while the dispersive configurations ($\Delta_1=\Delta_2 \gtrsim 5g$) achieve maximal fidelity over longer times (at $t_{f}^{'}(\Delta)$), and the intermediate configurations ($g < \Delta_1 = \Delta_2 < 5g$) achieve maximal fidelity over much longer times. Note the subtle asymmetry in darkness along the diagonal, between positive and negative equal detuning configurations (a non-RWA effect).

\section{The rotating wave approximation}
\label{sec:RWA}

\noindent The rotating wave approximation (RWA) is often performed in the analytical solution or in the numerical simulation of the Tavis-Cummings and parent Hamiltonians. It neglects the fast rotating terms in the interaction part of the Hamiltonian based on the following conditions:
\begin{eqnarray}
\Delta_i &<<& \omega_c + \omega_i \label{eq:RWAcondition1} \\
g &<<& \omega_c +\omega_i
\label{eq:RWAcondition2}
\end{eqnarray}
In this context, the Tavis-Cummings Hamiltonian in the RWA can be written as:
\begin{eqnarray}
H^{RWA} &=& \frac{\hbar \omega_1}{2}\sigma^z_1 + \frac{\hbar \omega_2}{2}\sigma^z_2 + \hbar \omega_c a^{\dagger}_c a_c \nonumber \\
&\quad& + \hbar g_1 (a_c^\dagger \sigma^-_1 + a_c \sigma^+_1) \nonumber \\ 
&\quad& + \hbar g_2 (a_c^\dagger \sigma^-_2 + a_c \sigma^+_2 ).
\end{eqnarray}
When this Hamiltonian is mapped onto qubits, with $\hbar=1$, we obtain the ``qubitized" RWA Hamiltonian (see Appendix \ref{sec:Appendix_Qubitization}):

\begin{eqnarray}
H_q^{RWA} &=& -\dfrac{\omega_1}{2}Z_1 -\dfrac{\omega_2}{2}Z_2 
+ \dfrac{\omega_c}{2}(I_c - Z_c) \nonumber \\ 
&\quad& + \dfrac{g_1}{2}(X_c X_1 + Y_c Y_1) \nonumber \\
&\quad& + \dfrac{g_2}{2}(X_c X_2 + Y_c Y_2).
\label{eq:qubitized_Hamiltonian_RWA}
\end{eqnarray}

\begin{figure}[t] 
    \centering
    \includegraphics[width=\columnwidth]{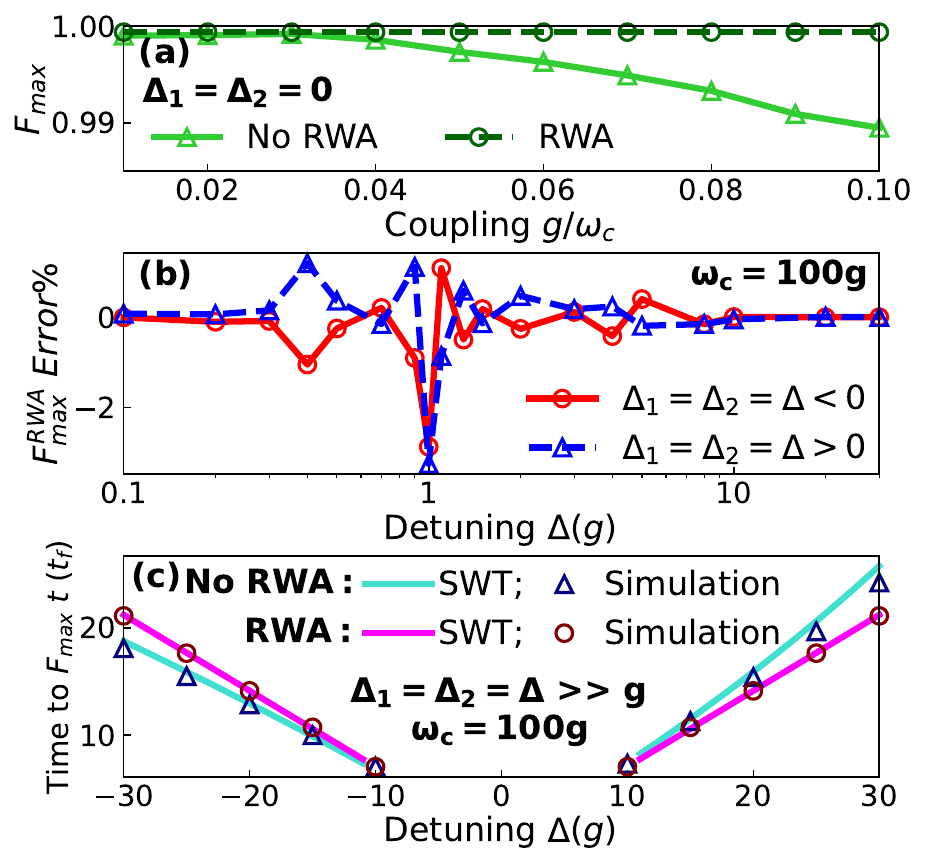}
    \caption{(a) Maximum achieved fidelity for the full Hamiltonian (No RWA) and the rotating wave approximation (RWA) Hamiltonian. (b) Error in maximum fidelity due to the RWA as a function of detunings in the case of 
    $\Delta_1 = \Delta_2 = \Delta$ for $\Delta$ positive (blue) and $\Delta$ negative (red). (c) Comparison of the time to maximum fidelity obtained through the quantum simulation with and without the rotating wave approximation versus the Schrieffer-Wolff transformation (SWT) result as a function of the detuning.}
    \label{fig:RWAvsNRWA}
\end{figure}

\noindent To characterize the effect of the rotating wave approximation, we consider the full Hamiltonian in the frame rotating at the cavity frequency for the case where $g_1=g_2=g$ (see appendix \ref{sec:Appendix_TavisCummings_CavityFrame}):
\begin{eqnarray}
H_{cavity} &=& \frac{\Delta_1}{2}\sigma^z_1 + \frac{\Delta_2}{2}\sigma^z_2 
+ g(a_c^\dagger \sigma^-_1 + a_c \sigma^+_1) \nonumber \\
&\quad& +  g(a_c^\dagger \sigma^-_2 + a_c \sigma^+_2) \nonumber \\
&\quad& + g e^{2i\omega_c t}(a_c^\dagger \sigma^+_1  + a_c^\dagger \sigma^+_2) \nonumber \\
&\quad& + g e^{-2i\omega_c t}(a_c \sigma^-_1 + a_c \sigma^-_2). 
\label{eqn:Tavis_Cummings_H_cavity}
\end{eqnarray}

\noindent We note that the non-energy-conserving terms (fifth and sixth terms) in Eq.~\ref{eqn:Tavis_Cummings_H_cavity} are fast oscillating and will average out over timescales corresponding to $\Delta$ or $g$ if the conditions in Eqs.~\ref{eq:RWAcondition1} and ~\ref{eq:RWAcondition2} are met.
We simulate both Hamiltonians using our quantum circuit approach. Figure \ref{fig:RWAvsNRWA} illustrates our comparison of the results from the full Hamiltonian against that of the RWA Hamiltonian when both Hamiltonians are simulated with the quantum algorithms. Fig.~\ref{fig:RWAvsNRWA} (a) compares the maximum achieved fidelities as a function of the cavity-qubit coupling strength for qubits resonant with the cavity. As the coupling strength increases, we note a deviation between the RWA and the full Hamiltonian result, as could be anticipated from the validity conditions of the RWA in Eqs.~\ref{eq:RWAcondition1} and ~\ref{eq:RWAcondition2}. Fig.~\ref{fig:RWAvsNRWA} (b) shows the error in the maximum fidelity due to the RWA as a function of the detuning in the case of $\Delta_1 = \Delta_2 = \Delta$ for $\Delta$ positive (blue) and $\Delta$ negative (red). The error in maximum fidelity obtained from the RWA simulation is  pronounced when $\Delta \sim g$, and shows an asymmetry with respect to $\Delta$ positive and $\Delta$ negative. Fig.~\ref{fig:RWAvsNRWA} (c) shows a comparison of the time to maximum fidelity obtained through the quantum simulation with and without RWA, alongside the analytical result using the Schrieffer-Wolff transformation in both cases as a function of the detuning.  It clearly indicates an asymmetry in the behaviors for $\Delta$ positive and for $\Delta$ negative. This asymmetry is clearly captured by the quantum simulation across the full range of the detuning values.
To understand its origin, consider the dispersive regime. Here, upon performing the Schrieffer-Wolff transformation, an effective Hamiltonian derived for the full Hamiltonian in Eq.~\ref{eq:Hamiltonian} can be obtained~\cite{Zueco2009_NRWA}. In this effective Hamiltonian, the effective interaction between the qubits acquires a correction that is indeed sensitive to the sign of detunings:
\begin{equation}
J = g_1 g_2 \bigg(\dfrac{1}{\Delta_1} + \dfrac{1}{\Delta_2} - \dfrac{1}{2 \omega_1 - \Delta_1} - \dfrac{1}{2 \omega_2 - \Delta_2}\bigg)
\label{eqn:Effective_Interaction_NRWA}
\end{equation}
In agreement with the quantum circuits, this results in an asymmetry of the dynamics between the $\Delta_1 = \Delta_2 = \Delta > 0$ case and the $\Delta_1 = \Delta_2 = \Delta < 0$ case. This is manifested in the measured time to achieve the first complete state transfer, which for equal detunings is:
\begin{eqnarray}
t_f^{''} = \dfrac{\pi}{g^2 (\frac{2}{\Delta} - \frac{2}{2 \omega_1 - \Delta})}
\label{eqn:Dispersive_Time_SWT}
\end{eqnarray}

\noindent It is important to remark here that in our results, the effect of the non-energy-conserving terms ($a^{\dagger}_c \sigma_i^+$ and $a_c \sigma_i^-$)  in the full Hamiltonian in Eq.~\ref{eq:Hamiltonian} is limited to the transitions $|e_1 0_c g_2 \rangle \leftrightarrow |e_1 1_c e_2 \rangle $ and $|g_1 0_c e_2 \rangle \leftrightarrow |e_1 1_c e_2 \rangle $ due to the capping of the cavity to a maximum of 1 excitation in our quantum circuit. Transitions to states with cavity occupation $>1$ are not captured, which implies that our results do not represent the complete extent of the deviation between the RWA and full Hamiltonian simulations. 
This limitation is addressed in Appendix \ref{sec:Appendix_Multilevel_Cavity}, where we evaluate state transfer fidelity as a function of coupling strength for the full Hamiltonian simulation with the cavity accommodating up to 3 excitations. As expected, with increasing coupling strength we observe a greater drop in fidelity relative to that seen in Figure \ref{fig:RWAvsNRWA}(a), due to the population of the higher excited states of the cavity.

\section{Conclusion}
\label{sec:conclusion}

\noindent We have implemented quantum circuits, that are compatible with NISQ (Noisy Intermediate Scale Quantum) era systems, to study the efficiency of quantum state transfer between two qubits placed in a cavity as a function of various system parameters. The quantum algorithm solves the Tavis-Cummings model describing the dynamics of the system, and allows us to reliably bypass conventional approximations that often inhibit analytical/numerical solutions, and to assess their regimes of validity. We are able to access the dispersive regime without resorting to a perturbative treatment such as the Schrieffer-Wolff transformation. We also solve the original Hamiltonian without performing the rotating wave approximation. Our approach allows us to map out the fidelity of the operation between the two qubits, and its dependence on various system parameters such as the emission frequencies of the qubits and the cavity frequency, as well as the respective qubit couplings to the cavity, and the damping factor of the cavity. Our quantum algorithm allows us to identify under-explored regimes of optimal performance where the two-qubit gate can be rather effective between far-detuned qubits that are neither resonant with each other nor with the cavity. We examine the effect of the rotating wave approximation on the state transfer operation. The robustness of the present solution offers a promising avenue for simulating and optimizing photon-mediated operations in quantum information processing beyond the cavity-mediated two-qubit gates examined in this paper.


\section*{Acknowledgment} 
\noindent We acknowledge support from the National Science Foundation under Grants No. PHY-2014023 and No. QIS-2328752. This work was also partially supported by AFRL/RI under contract number SA10032025060975. We thank W. A. Coish, V. V. Dobrovitski and T. Thomay for useful discussions.

\bibliography{References}


\onecolumngrid
\appendix
\newpage

\section{Mapping qubit and cavity operators to spin 1/2 operators in the single excitation subspace}
\label{sec:Appendix_HPTransformation}

\noindent Qubits live in the 2-dimensional state space of the spin 1/2 matrices. However, formally mapping them to the computational basis $\{\ket{0},\ket{1}\}$  needs to account for any basis convention differences between the system and the algorithm. In the computational basis, the ground and excited states are defined as

\begin{equation}
\ket{g} \equiv \ket{0} \equiv \begin{bmatrix}
1\\
0\\
\end{bmatrix}, \quad
\ket{e} \equiv \ket{1} \equiv \begin{bmatrix}
0\\
1\\
\end{bmatrix}
\end{equation}

\noindent The Pauli matrices corresponding to this basis are

\begin{equation}
X=\ket{e}\bra{g} + \ket{g}\bra{e}, \quad Y=i\ket{e}\bra{g} - i\ket{g}\bra{e}, \quad Z=\ket{g}\bra{g} - \ket{e}\bra{e}  
\end{equation}

\noindent Our Tavis-Cummings system is defined in the convention generally used in quantum optics, where the definition of $\ket{0}$ and $\ket{1}$ are flipped relative to that in the computational basis. This leads to the requirement of a formal mapping of our system's operators  operators to the computational Pauli matrices, which we perform below~\cite{Nielsen_Chuang_2010}.

\begin{eqnarray}
\sigma^z = \ket{e}\bra{e} - \ket{g}\bra{g} = -Z \\
\sigma^+ = \ket{e}\bra{g} = \dfrac{1}{2}(X - iY) \\
\sigma^- = \ket{g}\bra{e} = \dfrac{1}{2}(X + iY)
\end{eqnarray}

\noindent The bosonic cavity operators are defined as

\begin{eqnarray}
a &=& \sum_{n=1}^{\infty} \sqrt{n}\ket{n-1}\bra{n}  \\
a^\dagger &=& \sum_{n=0}^{\infty} \sqrt{n+1}\ket{n+1}\bra{n}
\end{eqnarray}

\noindent We map $a$ and $a^\dagger$ to the computational Pauli matrices by representing the cavity as a qubit. Such a representation is made possible by considering the system, initiated to have a maximum of 1 excitation, to be isolated. This restricts our state space to the single excitation subspace where $n=\{0,1\}$. The cavity can then be treated as a qubit, and its operators defined above can be  mapped to the computational Pauli matrices as 

\begin{eqnarray}
a = \ket{0}\bra{1}=\sigma^- = \dfrac{1}{2}(X+iY)\\ \quad a^\dagger = \ket{1}\bra{0} = \sigma^+ = \dfrac{1}{2}(X-iY)
\end{eqnarray}

\section{Qubitization of the Tavis-Cummings Hamiltonian}
\label{sec:Appendix_Qubitization}

\noindent Here, we "qubitize" the Tavis-Cummings Hamiltonian in Eq.~\ref{eq:Hamiltonian} 
using the operator transformations in Eq. \ref{eq:HP_transformations}. We first transform the interactions terms as follows.

\begin{eqnarray}
a^\dagger_c \sigma^-_i &=& \dfrac{1}{2}(X_c - iY_c)\frac{1}{2}(X_i + iY_i) = \dfrac{1}{4}(X_c X_i + iX_c Y_i - iY_c X_i + Y_c Y_i) \\
a_c \sigma^+_i &=& \dfrac{1}{2}(X_c + iY_c)\frac{1}{2}(X_i - iY_i) 
= \dfrac{1}{4}(X_c X_i - iX_c Y_i + iY_c X_i + Y_c Y_i) \\
a^\dagger_c \sigma^+_i &=& \dfrac{1}{2}(X_c - iY_c)\frac{1}{2}(X_i - iY_i) 
= \dfrac{1}{4}(X_c X_i - iX_c Y_i - iY_c X_i - Y_c Y_i) \label{eq:cr+} \\
a_c \sigma^-_i &=& \dfrac{1}{2}(X_c + iY_c)\frac{1}{2}(X_i + iY_i)
= \dfrac{1}{4}(X_c X_i + iX_c Y_i + iY_c X_i - Y_c Y_i)
\label{eq:cr-}
\end{eqnarray}

\noindent It is easy to see that all but the $X_c X_i$ terms cancel out in the above transformations, leading to a concise "qubitized" Hamiltonian, given by

\begin{eqnarray}
H_q &=& -\dfrac{\omega_1}{2}Z_1 -\dfrac{\omega_2}{2}Z_2 
+ \dfrac{\omega_c}{2}(I_c - Z_c)
+ g_1(X_c X_1) + g_2(X_c X_2)
\end{eqnarray}

\noindent When the rotating wave approximation is applied, we drop the non-energy-conserving terms (Eqs. \ref{eq:cr+} and  \ref{eq:cr-}) and obtain the "qubitized" Hamiltonian in the RWA as follows.

\begin{eqnarray}
H_q^{RWA} &=& -\dfrac{\omega_1}{2}Z_1 -\dfrac{\omega_2}{2}Z_2 
+ \dfrac{\omega_c}{2}(I_c - Z_c)
+ g_1(X_c X_1 + Y_c Y_1) + g_2(X_c X_2 + Y_c Y_2)
\end{eqnarray}

\section{Exact solution to the Tavis-Cummings Hamiltonian when the qubits are resonant with the cavity}
\label{sec:Appendix_ExactSolution}

\noindent In this section, we derive $\Psi(t)$ for the special case where the qubits are resonant with the cavity, and with an initial state of the system where \textit{Qubit 1} is in a superposition state, and the cavity and \textit{Qubit 2} are  empty and in the ground state respectively.

\begin{equation}
\ket{\Psi(0)} = \alpha \ket{g_1 0_c g_2} + \beta \ket{e_1 0_c g_2}
\end{equation}

\noindent The Tavis-Cummings Hamiltonian in the RWA then becomes

\begin{equation}
H = g_1(a^\dagger \sigma^-_1 + a \sigma^+_1) + g_2(a^\dagger \sigma^-_2 + a \sigma^+_2)
\end{equation}

\noindent Since we are interested in the time evolution of a superposition state, we consider the extended single excitation subspace accommodating the ground state component, given by

\begin{equation}
\mathcal{H}_{reduced} = \{ \ket{e_1 0_c g_2}, \ket{g_1 1_c g_2}, \ket{g_1 0_c e_2}, \ket{g_1 0_c g_2} \}
\end{equation}

\noindent In the subspace ordered as such, the matrix representation of the Hamiltonian is

\begin{equation}
H =
\begin{bmatrix}
0 & g_1 & 0 & 0 \\
g_1 & 0 & g_2 & 0 \\
0 & g_2 & 0 & 0 \\
0 & 0 & 0 & 0 \\
\end{bmatrix}
\end{equation}

\noindent We note that the state $\ket{g0g}$ is not affected by the Hamiltonian's dynamics, and is therefore preserved during the time evolution of the system. This, along with the linearity of the evolution operator $U(t)$, enables us to write

\begin{eqnarray}
\ket{\Psi(t)} &=& \alpha\ket{\Psi_g(t)} + \beta\ket{\Psi_e(t)} \\
&=& \alpha\ket{\Psi_g(0)} + \beta \bigg(U(t)\ket{\Psi_e(0)} \bigg) \\
&=& \alpha\ket{g_1 0_c g_2} + \beta \bigg( U(t)\ket{e_1 0_c g_2} \bigg)
\label{eqn:evolution_0}
\end{eqnarray}

\noindent We solve for $\Psi_e(t)$ using the method of exact diagonalization in the single excitation subspace $\{ \ket{e_1 0_c g_2}, \ket{g_1 1_c g_2}, \ket{g_1 0_c e_2} \}$. In this basis, the Hamiltonian in the matrix representation becomes

\begin{equation}
H =
\begin{bmatrix}
0 & g_1 & 0 \\
g_1 & 0 & g_2 \\
0 & g_2 & 0
\end{bmatrix}
\end{equation}

\noindent The eigenvalues and eigenvectors of H are:

\begin{eqnarray}
E_1 &=& 0\\
E_2 &=& \sqrt{g_1^2 + g_2^2}\\
E_3 &=& -\sqrt{g_1^2 + g_2^2}\\
\ket{E_1} &=& \dfrac{g_2}{\sqrt{g_1^2 + g_2^2}} \ket{e_1 0_c g_2} - \dfrac{g_1}{\sqrt{g_1^2 + g_2^2}} \ket{g_1 0_c e_2}\\
\ket{E_2} &=& \dfrac{1}{\sqrt{2(g_1^2 + g_2^2)}} \bigg(g_1 \ket{e_1 0_c g_2}  + g_2 \ket{g_1 0_c e_2} \bigg) + \dfrac{1}{\sqrt{2}}\ket{g_1 1_c g_2} \\
\ket{E_3} &=& \dfrac{1}{\sqrt{2(g_1^2 + g_2^2)}} \bigg(g_1 \ket{e_1 0_c g_2}  + g_2 \ket{g_1 0_c e_2} \bigg) - \dfrac{1}{\sqrt{2}}\ket{g_1 1_c g_2} \\
\end{eqnarray}

\noindent $\ket{e_1 0_c g_2}$ can be represented in the energy eigenbasis $\{\ket{E_i}\}$ as

\begin{equation}
\ket{e_1 0_c g_2} = \dfrac{g_2}{\sqrt{g_1^2 + g_2^2}} \ket{E_1} + \dfrac{g_1}{\sqrt{2(g_1^2 + g_2^2)}} \bigg( \ket{E_2} + \ket{E_3} \bigg)
\end{equation}

\noindent The evolution operator $U(t) = e^{-iHt}$ acts trivially in this representation.

\begin{eqnarray}
\ket{\Psi_e(t)} &=& U(t) \bigg( \dfrac{g_2}{\sqrt{g_1^2 + g_2^2}} \ket{E_1} + \dfrac{g_1}{\sqrt{2(g_1^2 + g_2^2)}} \bigg( \ket{E_2} + \ket{E_3} \bigg) \bigg) \\
&=& \dfrac{g_2}{\sqrt{g_1^2 + g_2^2}} \ket{E_1} + \dfrac{g_1}{\sqrt{2(g_1^2 + g_2^2)}} \bigg( e^{-i\sqrt{g_1^2 + g_2^2}t}\ket{E_2} + e^{i\sqrt{g_1^2 + g_2^2}t} \ket{E_3} \bigg)
\end{eqnarray}

\noindent Re-expressing in the original basis and grouping by basis states, we get

\begin{eqnarray}
\ket{\Psi_e(t)} &=& \bigg( \dfrac{g_2^2}{g_1^2 + g_2^2} 
+ \dfrac{(e^{-i\sqrt{g_1^2 + g_2^2}t} + e^{i\sqrt{g_1^2 + g_2^2}t}) g_1^2}{2(g_1^2 + g_2^2)} \bigg) \ket{e_1 0_c g_2} \nonumber \\
&\quad& + \bigg( \dfrac{(e^{-i\sqrt{g_1^2 + g_2^2}t} - e^{i\sqrt{g_1^2 + g_2^2}t}) g_1}{2\sqrt{g_1^2 + g_2^2}} \bigg) \ket{g_1 1_c g_2} \nonumber \\
&\quad& + \bigg( \dfrac{-g_1 g_2}{g_1^2 + g_2^2} + \dfrac{(e^{-i\sqrt{g_1^2 + g_2^2}t} + e^{i\sqrt{g_1^2 + g_2^2}t}) g_1 g_2}{2(g_1^2 + g_2^2)} \bigg) \ket{g_1 0_c e_2}
\end{eqnarray}

\begin{eqnarray}
\ket{\Psi_e(t)} &=& \dfrac{g_1^2 cos \bigg(\sqrt{g_1^2 + g_2^2}t \bigg) + g_2^2}{g_1^2 + g_2^2} \ket{e_1 0_c g_2}
- \dfrac{i g_1 sin\bigg(\sqrt{g_1^2 + g_2^2}t \bigg)}{\sqrt{g_1^2 + g_2^2}} \ket{g_1 1_c g_2} \nonumber \\ 
&\quad& + \dfrac{g_1g_2 \bigg( cos \bigg(\sqrt{g_1^2 + g_2^2}t \bigg) - 1 \bigg)}{g_1^2 + g_2^2} \ket{g_1 0_c e_2}
\end{eqnarray}

\noindent Plugging this into Eq. \ref{eqn:evolution_0}, we get the exact solution for a superposition state as:

\begin{eqnarray}
\ket{\Psi(t)} &=& \alpha \ket{g_1 0_c g_2} + \beta \bigg[ \dfrac{g_1^2 cos \bigg(\sqrt{g_1^2 + g_2^2}t \bigg) + g_2^2}{g_1^2 + g_2^2} \ket{e_1 0_c g_2}
- \dfrac{i g_1 sin\bigg(\sqrt{g_1^2 + g_2^2}t \bigg)}{\sqrt{g_1^2 + g_2^2}} \ket{g_1 1_c g_2} \nonumber \\
&\quad& \quad\quad\quad\quad\quad\quad + \dfrac{g_1g_2 \bigg( cos \bigg(\sqrt{g_1^2 + g_2^2}t \bigg) - 1 \bigg)}{g_1^2 + g_2^2} \ket{g_1 0_c e_2} \bigg]
\label{eqn:Psi(t)_Delta_0_g_unequal}
\end{eqnarray}

\noindent For $g_1 = g_2$, we get

\begin{eqnarray}
\ket{\Psi(t)} &=& \alpha \ket{g_1 0_c g_2} + \beta \bigg[ \dfrac{1 + cos \big(\sqrt{2}gt \big)}{2} \ket{e_1 0_c g_2} - \dfrac{i sin \big(\sqrt{2}gt \big)}{\sqrt{2}} \ket{g_1 1_c g_2} 
+ \dfrac{cos \big(\sqrt{2}gt \big) - 1}{2}  \ket{g_1 0_c e_2} \bigg]
\label{eqn:Psi(t)_Delta_0_g_equal}
\end{eqnarray}

\noindent At $t=0$, we recover the initial state.
\begin{equation}
\ket{\Psi(0)} = \alpha\ket{g_1 0_c g_2} + \beta\ket{e_1 0_c g_2}   
\end{equation}

\noindent At $t=t_f=\pi/2\sqrt{2}g$, the cavity reaches its maximum occupation, which is equal to half the total excitation.
\begin{equation}
\ket{\Psi(\pi/2\sqrt{2}g)} = \alpha \ket{g_1 0_c g_2} + \beta \bigg( \dfrac{1}{2} \ket{e_1 0_c g_2}
- \dfrac{i}{\sqrt{2}} \ket{g_1 1_c g_2}
- \dfrac{1}{2} \ket{g_1 0_c e_2} \bigg)
\end{equation}
 
\noindent At $t=\pi/\sqrt{2}g$, there is complete state transfer from Qubit 1 to Qubit 2, with the relative phase shifted by $\pi$.
\begin{equation}
\ket{\Psi(\pi/\sqrt{2}g)} = \alpha\ket{g_1 0_c g_2} - \beta\ket{g_1 0_c e_2}
\label{eqn:time_for_transfer}
\end{equation}

\noindent This phase shift can be corrected  using a $Z$ gate.

\section{The Tavis-Cummings Hamiltonian in the cavity frame}
\label{sec:Appendix_TavisCummings_CavityFrame}

\noindent In order to express the Hamiltonian only in terms of the detunings $\Delta_i$ and the couplings $g_i$, we move $H$ in Eq. \ref{eq:Hamiltonian} to the rotating frame of the cavity $\omega_c$ using a frame transformation. 

\begin{equation}
    H_{cavity} = U H U^\dagger - iU \dfrac{\partial U^\dagger}{\partial t}
\label{eqn:frame_transformation_general}
\end{equation}

\noindent where $U(t)=e^{iGt}$ is the frame rotation operator based on the generator

\begin{equation}
    G = \omega_c \bigg(\dfrac{\sigma^z_1 }{2}+ \dfrac{\sigma^z_2 }{2} + a^\dagger_c a_c \bigg)
\end{equation}

\noindent Since $G$ is time independent, Eq. \ref{eqn:frame_transformation_general} reduces to

\begin{eqnarray}
H_{cavity} &=& U H U^\dagger - iU (-iG)U^\dagger \nonumber \\
&=& UH U^\dagger - G \nonumber \\
&=& U H U^\dagger - \omega_c \bigg(\dfrac{\sigma^z_1 }{2}+ \dfrac{\sigma^z_2 }{2} + a^\dagger_c a_c \bigg) 
\label{eqn:H_transformation}
\end{eqnarray}

\noindent Using the relation $e^{i\theta G}Ae^{i\theta G} = e^{i\theta \lambda}A$ when $[G,A] = -\lambda A$, we get

\begin{eqnarray}
U\sigma^z_iU^\dagger &=& \sigma^z_i \nonumber \\
U\sigma^+_iU^\dagger &=& e^{i \omega_c t}\sigma^+_i \nonumber \\
U\sigma^-_iU^\dagger &=& e^{-i \omega_c t}\sigma^-_i \nonumber \\
U a^\dagger_c U^\dagger &=& e^{i \omega_c t}a^\dagger_c \nonumber \\
U a_c U^\dagger &=& e^{-i \omega
_c t}a_c
\label{eqn:operator_rotations}
\end{eqnarray}

\noindent Plugging the operator transformations from Eqs. \ref{eqn:operator_rotations} into the Hamiltonian transformation in Eq. \ref{eqn:H_transformation}, we get the 2-qubit Tavis Cummings Hamiltonian in the rotating frame of the cavity, with $\hbar=1$ as follows.

\begin{equation}
H_{cavity} = \frac{\Delta_1}{2}\sigma^z_1 + \frac{\Delta_2}{2}\sigma^z_2 
+ g_1 (a_c^\dagger \sigma^-_1 + a_c \sigma^+_1) + g_2(a_c^\dagger \sigma^-_2 + a_c \sigma^+_2)
+ e^{2i\omega_c t} (g_1a_c^\dagger \sigma^+_1  + g_2 a_c^\dagger \sigma^+_2)
+ e^{-2i\omega_c t}(g_1 a_c \sigma^-_1 + g_2a_c \sigma^-_2) 
\label{eqn:Tavis_Cummings_H_cavity_frame}
\end{equation}

\noindent We note that rotation to the cavity frame introduces fast oscillations to the couplings of the non-energy-conserving terms.

\section{Estimating the systematic error in the simulation of the "qubitized" Hamiltonian using the cavity frame}
\label{sec:Appendix_Trotter_Error}

\noindent The systematic Trotter error is typically associated with the time scales (inverse frequencies) involved in the system. However, there is a risk of overestimating this error if we do not recognize the relative dynamics in the system. In our Tavis-Cummings Hamiltonian, we note that the relative dynamics are of the order $g,\Delta$ which can be seen by transforming the Hamiltonian to the cavity frame as $H_{cavity}$ (Eq. \ref{eqn:Tavis_Cummings_H_cavity_frame}). We proceed to ``qubitize" $H_{cavity}$ using the relations in Eq. \ref{eq:HP_transformations} and get

\begin{eqnarray}
H_{cavity}^q &=& \sum_{i=1,2}   -\dfrac{\Delta_i}{2}Z_i + \dfrac{g_i}{2}(X_c X_i + Y_c Y_i) 
+ \dfrac{g_i}{4}e^{2i\omega_c t}(X_c X_i - Y_c Y_i - iX_cY_i - iY_cX_i) \nonumber \\
&\quad& \quad \quad + \dfrac{g_i}{4}e^{-2i\omega_c t}(X_c X_i - Y_c Y_i + iX_cY_i + iY_cX_i) \nonumber \\
&=& \sum_{i=1,2}   -\dfrac{\Delta_i}{2}Z_i + \dfrac{g_i}{2}(X_c X_i + Y_c Y_i) 
+ \dfrac{g_i}{4}(e^{2i\omega_c t} + e^{-2i\omega_c t}) (X_c X_i - Y_c Y_i) 
- i\dfrac{g_i}{4}(e^{2i\omega_c t} - e^{-2i\omega_c t})(X_cY_i + iY_cX_i) \nonumber \\
&=& \sum_{i=1,2}   -\dfrac{\Delta_i}{2}Z_i + \dfrac{g_i}{2}(X_c X_i + Y_c Y_i) 
+ \dfrac{g_i}{2}cos(2 \omega_c t) (X_c X_i - Y_c Y_i) 
+ \dfrac{g_i}{2}sin(2 \omega_c t)(X_cY_i + iY_cX_i)
\end{eqnarray}

\noindent We note that $H_{cavity}^q$ has the structure $H_i^0 + H_{ci}$ where $H_i^0$ is the free evolution of the $i^{th}$ qubit and $H_{ci}$ is the interaction of the cavity with the $i^{th}$ qubit. We leverage this structure to split the Hamiltonian as follows.

\begin{equation}
H_{cavity}^q=H^0 + H_{c1} + H_{c2}
\end{equation}

\noindent where $H^0=H_1^0+H_2^0$. The first order trotter error per step, $\epsilon = ||e^{-iH\delta t}-U_{ST}(\delta t)||$, is estimated as:

\begin{eqnarray}
\epsilon &\sim& \dfrac{\delta t^2}{2} \bigg( [H^0,H_{1c}] + [H^0,H_{2c}]  + [H_{1c},H_{2c}] \bigg) \nonumber \\
&\sim& \delta t^2 \bigg( g_1 \Delta_1  + g_2 \Delta_2  + g_1 g_2 \bigg)
\end{eqnarray}

The total error over a simulation time $T$, represented by $N=T/\delta t$ steps, accumulates at most linearly \cite{Layden2022_TrotterError}, and therefore has an upper bound given by

\begin{equation}
\epsilon_{tot} \sim \bigg( g_1 \Delta_1  + g_2 \Delta_2  + g_1 g_2 \bigg) T \delta t
\end{equation}
\label{eqn:Trotter_Error_Total}

\noindent This conservative estimate serves to provide a reasonable reference time step size for the corresponding simulation, which can be further optimized through empirical methods.

\section{Quantum Circuit for the Tavis-Cummings Hamiltonian with a damped cavity}
\label{sec:Appendix_Damped_Cavity}

\begin{figure}[t]
    \centering
    \includegraphics[width=12cm]{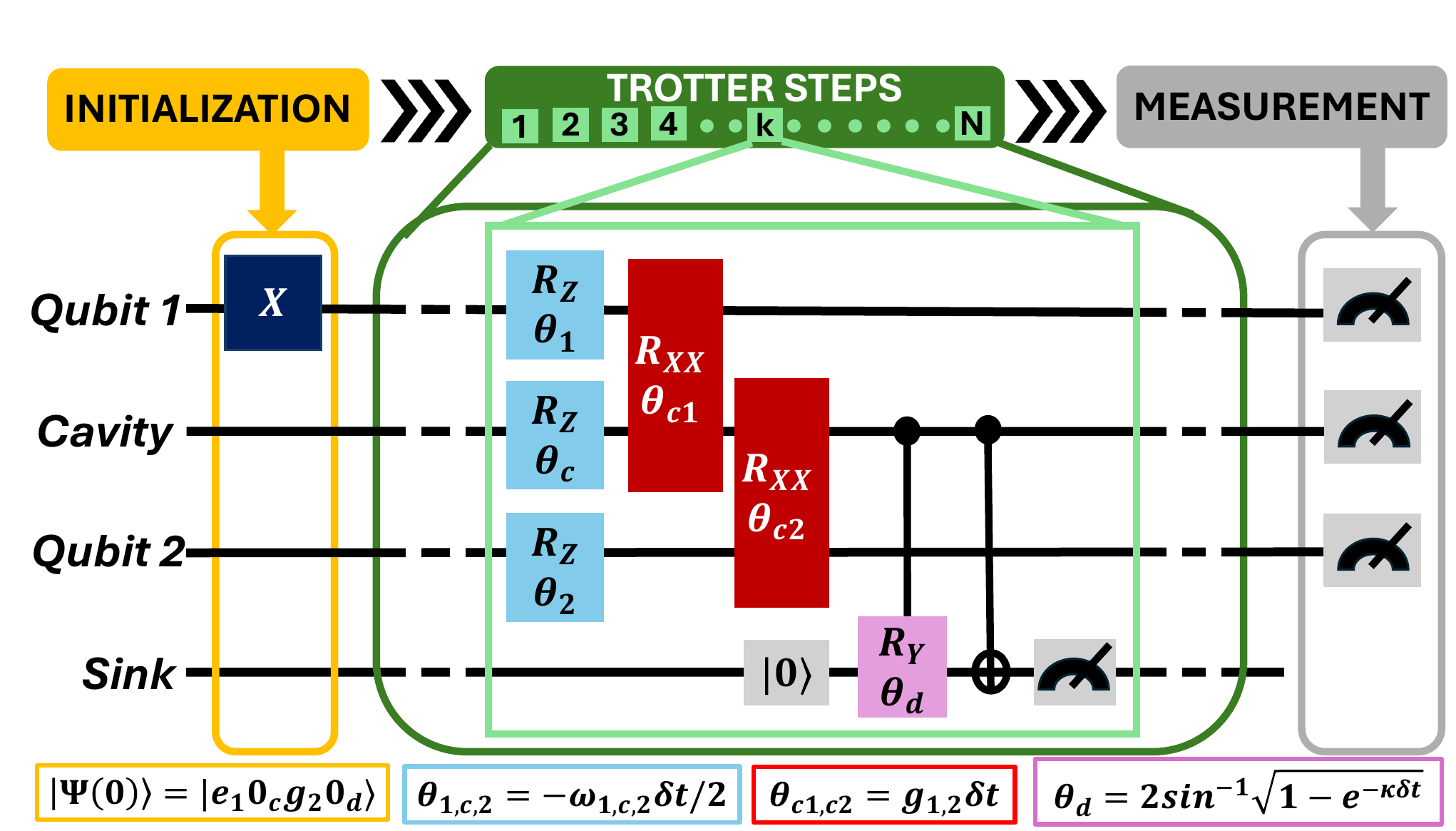}
    \caption{Quantum circuit for simulating the Tavis-Cummings Hamiltonian along with cavity damping. The arrangement involves a ancillary damping qubit called the sink which draws population out of the cavity at a rate $\kappa$. The sink is reset to $\ket{0}$ before each Trotter step and measured after it, to ensure a ``one-way” flow of excitation to it from the cavity.}
\label{fig:Quantum_Circuit_Damped}
\end{figure}

\noindent We incorporate cavity damping in our Tavis-Cummings quantum circuit using an amplitude damping circuit developed by Gupta et al. \cite{gupta2020optimalquantumsimulationopen}. We append their circuit to the qubit that represents the cavity in our circuit, as shown in Fig. \ref{fig:Quantum_Circuit_Damped}. The arrangement is essentially a damping channel $\mathcal{M}$ based on Kraus operators that act on the cavity's density matrix $\rho_c$ after every Trotter step of the Hamiltonian simulation.

\begin{equation}
\mathcal{M}(\rho_c) = \Omega_0 \rho_c \Omega_0^\dagger + \Omega_1 \rho_c \Omega_1^\dagger
\end{equation}

\noindent $\Omega_i$ are the Kraus operators defined as

\begin{eqnarray}
\Omega_0 &=& \ket{0}\bra{0} + \sqrt{1 - \gamma^2} \ket{1}\bra{1} \\
\Omega_1 &=& \gamma \ket{0}\bra{1}
\end{eqnarray}

\noindent $\gamma$ is the damping amplitude which we relate to the cavity damping factor $\kappa$ as

\begin{eqnarray}
\gamma = \sqrt{1-e^{-\kappa \delta t}}
\end{eqnarray}

\noindent In their paper, Gupta et. al show that the rotation angle $\theta_d$ that produces the effect of this damping channel over a Trotter step is given by

\begin{equation}
\theta_d = 2sin^{-1} \gamma = 2sin^{-1}\sqrt{1-e^{-\kappa \delta t}}
\end{equation}

\section{Evaluating the Rotating Wave Approximation via a 4-level cavity}
\label{sec:Appendix_Multilevel_Cavity}

\noindent In our model of the Tavis-Cummings Hamiltonian in Eq. 1, we treat the cavity as a two-level system by reasoning that the system can be considered to be isolated during state transfer timescales, effectively capping the total excitation in the system to its initial excitation, which we set to a maximum of one. However, the non-energy-conserving terms in the Hamiltonian, $a^\dagger_c \sigma^+_i$ and $a_c \sigma^-_i$, can produce transient virtual transitions to higher energy states. By capping the cavity at one excitation, the only state with a total excitation $n > 1$ that is accessible to the Hamiltonian is $|e_1 1_c e_2\rangle$ through the transitions $|e_1 0_c g_2\rangle \leftrightarrow |e_1 1_c e_2\rangle$ and $|g_1 0_c e_2\rangle \leftrightarrow |e_1 1_c e_2\rangle$. Since a cavity mode, in truth, can accommodate any number of excitations, the possibility of virtual transitions to states with higher cavity occupation (eg. $|e_1 2_c g_2\rangle$)  must be evaluated appropriately. In section \ref{sec:RWA}, we argued that the effect of the non-energy-conserving terms is negligible within the regime of validity of the RWA, but becomes significant away from thQuantum circuit for the time evolution of the ini-is regime. We recognize that the extent of this effect is understated with the cavity occupation limited to a maximum of one, and proceed to address this limitation by modeling the cavity with an occupation capacity of 3 photons through an efficient mapping onto just two qubits using the binary representation \cite{peng2023quantumsimulationbosonrelatedhamiltonians}. The qubitized Hamiltonian of the system with this 4-level cavity is given by

\begin{eqnarray}
    H_q &=& -\dfrac{\omega_1}{2}Z_1 -\dfrac{\omega_2}{2}Z_2 + \dfrac{\omega_c}{2}\bigg(\dfrac{3}{2}I_{c1}I_{c2} - Z_{c1}I_{c2} - \dfrac{1}{2}I_{c1}Z_{c2}\bigg) 
    \nonumber \\ &\quad\quad& + g_1 \bigg(\dfrac{1+\sqrt{3}}{2}X_1 I_{c1} X_{c2} + \dfrac{1-\sqrt{3}}{2}X_1  Z_{c1} X_{c2}) + \dfrac{1}{\sqrt{2}}X_1 X_{c1} X_{c2} + \dfrac{1}{\sqrt{2}}X_1 Y_{c1} Y_{c2}\bigg) \nonumber \\
    &\quad\quad& + g_2 \bigg(\dfrac{1+\sqrt{3}}{2}X_{2} I_{c1} X_{c2} + \dfrac{1-\sqrt{3}}{2} X_{2} Z_{c1} X_{c2}) + \dfrac{1}{\sqrt{2}} X_{2} X_{c1} X_{c2} + \dfrac{1}{\sqrt{2}}X_{2} Y_{c1} Y_{c2}\bigg)
\end{eqnarray}

\begin{figure}[t] 
    \centering
    \includegraphics[width=12.0cm]{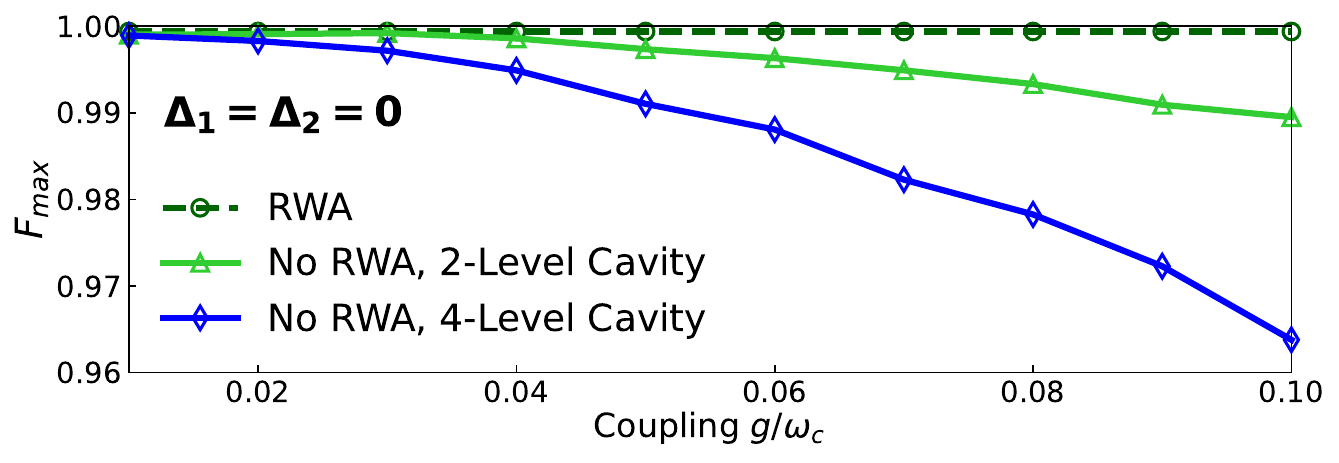}
    \caption{Maximum achieved fidelity as a function of coupling strength, for the transfer of the polarized state when the qubits are resonant with the cavity, using the Rotating Wave Approximation Hamiltonian (RWA), the full Hamiltonian with a 2-level cavity (No RWA, 2-Level Cavity) and the full Hamiltonian with a 4-level cavity (No RWA, 4-Level Cavity).}
    \label{fig:fidelity_multilevel_cavity}
\end{figure}

\noindent Figure \ref{fig:fidelity_multilevel_cavity} shows the state transfer fidelity for the resonant case ($\Delta_1=\Delta_2=0$) as a function of coupling strength, for simulations with the RWA Hamiltonian, the full Hamiltonian with a 2-level cavity, and the full Hamiltonian with a 4-level cavity. As expected, the effect of the non-energy-conserving terms is more pronounced when the cavity accommodates higher Fock states, opening up new transition pathways. As we enter the ultra-strong coupling regime ($g/\omega_c = 0.1$), there is a drop in fidelity of $\sim 4\%$ for the case of the  4-level cavity, compared to the $\sim 1\%$ drop in the case of the 2-level cavity. \\

\begin{figure}[t] 
    \centering
    \includegraphics[width=12.0cm]{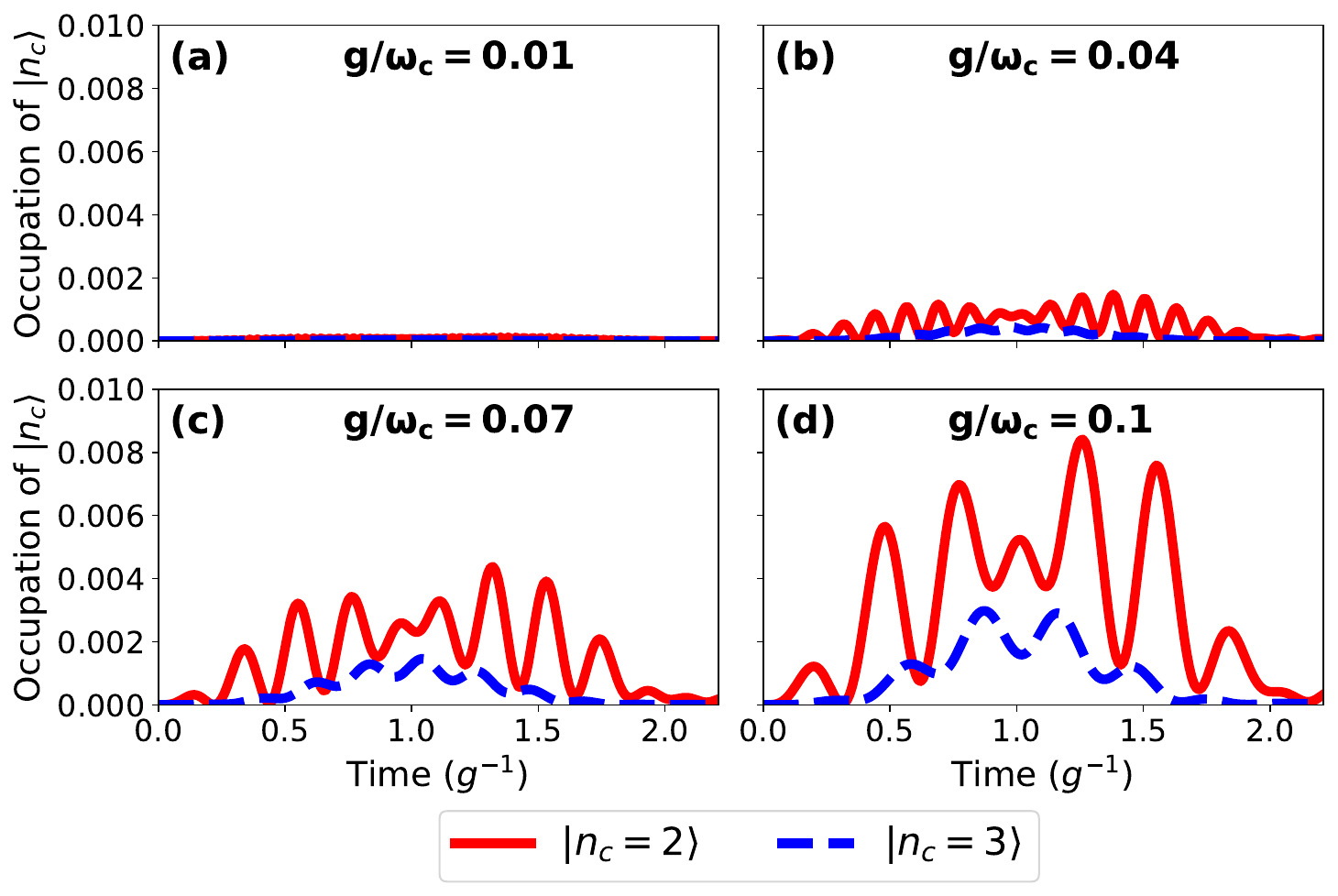}
    \caption{Occupation of the second and third excited states, $|n_c=2\rangle$ and $|n_c=3\rangle$ respectively, of the cavity as a function of time during the transfer of the polarized state when the qubits are resonant with the cavity, for a range of coupling strengths $g/\omega_c$.}
    \label{fig:multilevel_cavity_occupation}
\end{figure}

\noindent With just 1 excitation assigned to the isolated system initially, we expect that the occupation of even higher excited states ($n_c>3$) of the cavity through the effect of the non-energy-conserving terms would be negligible for the range of coupling strengths in Figure \ref{fig:fidelity_multilevel_cavity}. This is confirmed in Figure \ref{fig:multilevel_cavity_occupation} which shows the occupancy of the second and third excited states of the cavity as a function of time, for a range of coupling strengths. As expected, the occupation of these excited states picks up as the coupling strength increases. At ultra-strong coupling ($g/\omega_c=0.1$), where we see the maximum occupation of these excited states, we note that the maximum occupation of $|n_c=2\rangle$ contributes to $\sim 0.85\%$ of the total cavity occupation, whereas the maximum occupation of $|n_c=3\rangle$ contributes only $\sim 0.3\%$. The population of the higher excited states ($n_c>3$) of the cavity can be reasonably considered to be negligible. 

\end{document}